\documentclass[
    10pt,
    twocolumn,
    ]{article}

\usepackage{widetext}

\usepackage[utf8]{inputenc}
\usepackage[
colorlinks=true,
linkcolor=blue,
citecolor=blue,
filecolor=blue,
urlcolor=black,
plainpages=false,
pdfpagelabels,
breaklinks=true,
pdfusetitle,
]{hyperref}

\usepackage{authblk}

\usepackage[
style=numeric,
citestyle=numeric-comp,
sorting=none,
giveninits=true,
isbn=false,
doi=true,
uniquename=init,
]{biblatex}
    \renewbibmacro*{doi+eprint+url}{%
        \iftoggle{bbx:url}
        {\iffieldundef{doi}{\usebibmacro{url+urldate}}{}}
        {}%
        \newunit\newblock%
        \iftoggle{bbx:eprint}
        {\usebibmacro{eprint}}
        {}%
        \newunit\newblock%
        \iftoggle{bbx:doi}
        {\printfield{doi}}
        {}
    }
    \DeclareFieldFormat{doi}{%
        \mkbibacro{DOI}\addcolon\space
        \ifhyperref
            {\href{https://doi.org/#1}{\nolinkurl{#1}}}
        {\nolinkurl{#1}}}

\usepackage{algorithm}
\usepackage{algpseudocode}
\usepackage{amsthm, amsmath, amsfonts, amssymb, amscd}
\usepackage{bbm}
\usepackage{bm}
\usepackage{bookmark}
\usepackage{booktabs}
\usepackage[font=footnotesize, labelfont=bf]{caption}

\usepackage{cleveref}
\usepackage{enumitem}
\usepackage[a4paper, margin=20pt]{geometry}
\usepackage[docdef=atom]{glossaries-extra}
\usepackage{graphicx} 
\usepackage{lineno}
\usepackage{listings}
\usepackage{makecell}
\usepackage{mathtools}
\usepackage{parskip}
\usepackage{placeins} 
\usepackage{relsize}
\usepackage{setspace}
\usepackage{siunitx}
\usepackage{tabularx}

\usepackage{xcolor}
\usepackage{xparse}
\usepackage{xspace}

\usepackage{tikz}
\usepackage{dsfont}

\bookmarksetup{depth=4}
\definecolor{lightgrey}{gray}{0.9}
\crefname{equation}{Eqn.}{Eqns.}
\Crefname{equation}{Eqn.}{Eqns.}
\crefname{figure}{Fig.}{Figs.}
\Crefname{figure}{Fig.}{Figs.}
\crefname{table}{Table}{Tables}
\Crefname{table}{Table}{Tables}
\crefname{algorithm}{Algorithm}{Algorithms}
\Crefname{algorithm}{Algorithm}{Algorithms}

\setabbreviationstyle[acronym]{long-short}
\glssetcategoryattribute{acronym}{nohyperfirst}{true}

\newacronym{ad}{AD}{automatic differentiation}
	
\newacronym{am}{AM}{adjoint method}
	\newcommand{\am}{\gls{am}\xspace}
	
\newacronym{ann}{ANN}{artificial neural network}

\newacronym{asr}{ASR}{automatic speech recognition}
\newacronym{bptt}{BPTT}{backpropagation through time}
	\newcommand{\bptt}{\gls{bptt}\xspace}
\newacronym{ce}{CE}{cross-entropy}
\newacronym{cifar10}{CIFAR-10}{CIFAR-10}
	
\newacronym{cmc}{CMC}{cortical microcircuit}

\newacronym{cnn}{CNN}{convolutional neural network}

\newacronym{cvnn}{CVNN}{complex-valued neural network}

\newacronym{dl}{DL}{deep learning}
	
\newacronym{esn}{ESN}{echo state network}
\newacronym{gal}{GAL}{gain alignment learning}
	
\newacronym{gd}{GD}{gradient descent}
	
\newacronym{gle}{GLE}{Generalized Latent Equilibrium}
	\newcommand{\gle}{\gls{gle}\xspace}
\newacronym{gru}{GRU}{gated recurrent unit}

\newacronym{gsc}{GSC}{Google Speech Commands}
	
\newacronym{ie}{IE}{instantaneous errors}
\newacronym{le}{LE}{Latent Equilibrium}
	\renewcommand{\le}{\gls{le}\xspace}
\newacronym{li}{LI}{leaky integrator}
\newacronym{lif}{LIF}{leaky integrate-and-fire}
\newacronym{lrnn}{LRNN}{locally recurrent neural network}

\newacronym{lru}{LRU}{linear recurrent unit}

\newacronym{lstm}{LSTM}{long short-term memory}
	
\newacronym{lti}{LTI}{linear time-invariant}
	
\newacronym{mfcc}{MFCC}{Mel-frequency cepstral coefficient}
\newacronym{mfs}{MFS}{Mel-frequency spectrogram}
\newacronym{ml}{ML}{machine learning}
	\newcommand{\ml}{machine learning\xspace}
\newacronym{mlp}{MLP}{multi-layer perceptron}
	
\newacronym{mnist1d}{MNIST-1D}{MNIST-1D}
	\newcommand{\mnistId}{MNIST-1D\xspace}
\newacronym{mse}{MSE}{mean-squared error}
\newacronym{nla}{NLA}{neuronal least action}
	\newcommand{\nla}{\gls{nla}\xspace}
\newacronym{nlif}{nLIF}{non-leaky integrate-and-fire}
\newacronym{ode}{ODE}{ordinary differential equation}
\newacronym{pal}{PAL}{phaseless alignment learning}
	\newcommand{\pal}{\gls{pal}\xspace}
\newacronym{pyr}{PYR}{pyramidal}
	
\newacronym{rflo}{RFLO}{random feedback online learning}
	\newcommand{\rflo}{\gls{rflo}\xspace}
\newacronym{rl}{RL}{reinforcement learning}
\newacronym{rnn}{RNN}{recurrent neural network}

\newacronym{rtrl}{RTRL}{real-time recurrent learning}
	\newcommand{\rtrl}{\gls{rtrl}\xspace}
\newacronym{sgd}{SGD}{stochastic gradient descent}
	
\newacronym{snn}{SNN}{spiking neural network}
\newacronym{ssm}{SSM}{state space model}

\newacronym{sst}{SST}{somatostatin-expressing}
	
\newacronym{tcn}{TCN}{temporal convolutional network}

\newacronym{tle}{TLE}{temporal latent equilibrium}
	
\newacronym{vle}{VLE}{variational latent equilibrium}
	\newcommand{\vle}{\gls{vle}\xspace}
\newacronym{ttfs}{TTFS}{time-to-first-spike}

\newcommand{\inlinecode}[2]{ \colorbox{lightgrey}{ \lstinline[language=#1]{$#2$}} } %[]
\renewbibmacro*{name:andothers}{
    \ifboolexpr{
    test {\ifnumequal{\value{listcount}}{\value{liststop}}}
    and
    test \ifmorenames 
  }
    {\ifnumgreater{\value{liststop}}{1}
       {\finalandcomma}
       {}%
     \andothersdelim\bibstring{andothers}}
 {}
}

\newcommand{\AND}{\text{and}}
\newcommand{\au}{\text{a.u.}}
\newcommand{\barm}[1]{\overline{#1}^{\text{m}}}
\newcommand{\barr}[1]{\overline{#1}^{\text{r}}}
\newcommand{\Bell}{\bm b_\ell}
\newcommand{\bg}{B^\mathrm{G}}
\newcommand{\bgdot}{\dot B^\mathrm{G}}
\newcommand{\bigO}{\mathcal O}
\newcommand{\bp}{B^\mathrm{P}}
\newcommand{\bpdot}{\dot B^\mathrm{P}}
\newcommand{\brevem}[1]{\breve{#1}^{\text{m}}}
\newcommand{\brever}[1]{\breve{#1}^{\text{r}}}
\newcommand{\bs}[1]{\boldsymbol{#1}}
\newcommand{\btau}{\bs{\tau}}
\newcommand{\btaum}{\bs{\tau}^\mathrm{m}}
\newcommand{\btaumell}{\bs{\tau}^\mathrm{m}_\ell}
\newcommand{\btaur}{\bs{\tau}^\mathrm{r}}
\newcommand{\btaurell}{\bs{\tau}^\mathrm{r}_\ell}
\newcommand{\dd}{\text d}

\NewDocumentCommand{\dminus}{o}{
	\IfNoValueTF{#1}
	{\op{\mathcal D}{-}{\tau}}
	{\op{\mathcal D}{-}{\tau}[#1]}
}
\newcommand{\dminushat}{\op{\widehat{\mathcal D}}{-}{\tau}}
\NewDocumentCommand{\dminusmi}{o}{
	\IfNoValueTF{#1}
	{\op{\mathcal D}{-}{\taum_i}}
	{\op{\mathcal D}{-}{\taum_i}[#1]}
}
\NewDocumentCommand{\dminusm}{o}{
	\IfNoValueTF{#1}
	{\op{\mathcal D}{-}{\taum}}
	{\op{\mathcal D}{-}{\taum}[#1]}
}
\NewDocumentCommand{\Dminusm}{o}{
	\IfNoValueTF{#1}
	{\op{\bs{\mathcal D}}{-}{\btaum}}
	{\op{\bs{\mathcal D}}{-}{\btaum}[#1]}
}
\NewDocumentCommand{\Dminusr}{o}{
	\IfNoValueTF{#1}
	{\op{\bs{\mathcal D}}{-}{\btaur}}
	{\op{\bs{\mathcal D}}{-}{\btaur}[#1]}
}
\NewDocumentCommand{\Dminusrell}{o}{
	\IfNoValueTF{#1}
	{\op{\bs{\mathcal D}}{-}{\btaurell}}
	{\op{\bs{\mathcal D}}{-}{\btaurell}[#1]}
}
\NewDocumentCommand{\DminusrL}{o}{
	\IfNoValueTF{#1}
	{\op{\bs{\mathcal D}}{-}{\btaur_L}}
	{\op{\bs{\mathcal D}}{-}{\btaur_L}[#1]}
}
\NewDocumentCommand{\dminusr}{o}{
	\IfNoValueTF{#1}
	{\op{\mathcal D}{-}{\taur}}
	{\op{\mathcal D}{-}{\taur}[#1]}
}
\NewDocumentCommand{\dminusri}{o}{
	\IfNoValueTF{#1}
	{\op{\mathcal D}{-}{\taur_i}}
	{\op{\mathcal D}{-}{\taur_i}[#1]}
}
\NewDocumentCommand{\dminusrj}{o}{
	\IfNoValueTF{#1}
	{\op{\mathcal D}{-}{\taur_j}}
	{\op{\mathcal D}{-}{\taur_j}[#1]}
}
\newcommand{\dminusrhat}{\op{\widehat{\mathcal D}}{-}{\taur}}
\NewDocumentCommand{\dminusrell}{o}{
	\IfNoValueTF{#1}
	{\op{\mathcal D}{-}{\taurell}}
	{\op{\mathcal D}{-}{\taurell}[#1]}
}
\NewDocumentCommand{\dplusrell}{o}{
	\IfNoValueTF{#1}
	{\op{\mathcal D}{+}{\taurell}}
	{\op{\mathcal D}{+}{\taurell}[#1]}
}
\NewDocumentCommand{\dplusri}{o}{
	\IfNoValueTF{#1}
	{\op{\mathcal D}{+}{\taur_i}}
	{\op{\mathcal D}{+}{\taur_i}[#1]}
}
\NewDocumentCommand{\dplusrj}{o}{
	\IfNoValueTF{#1}
	{\op{\mathcal D}{+}{\taur_j}}
	{\op{\mathcal D}{+}{\taur_j}[#1]}
}
\NewDocumentCommand{\dplus}{o}{
	\IfNoValueTF{#1}
	{\op{\mathcal D}{+}{\tau}}
	{\op{\mathcal D}{+}{\tau}[#1]}
}
\newcommand{\dplushat}{\op{\widehat{\mathcal D}}{+}{\tau}}

\NewDocumentCommand{\dplusm}{o}{
	\IfNoValueTF{#1}
	{\op{\mathcal D}{+}{\taum}}
	{\op{\mathcal D}{+}{\taum}[#1]}
}
\newcommand{\dplusmhat}{\op{\widehat{\mathcal D}}{+}{\taum}}
\NewDocumentCommand{\Dplusm}{o}{
	\IfNoValueTF{#1}
	{\op{\bs{\mathcal D}}{+}{\btaum}}
	{\op{\bs{\mathcal D}}{+}{\btaum}[#1]}
}
\NewDocumentCommand{\Dplus}{o}{
	\IfNoValueTF{#1}
	{\op{\bs{\mathcal D}}{+}{\btau}}
	{\op{\bs{\mathcal D}}{+}{\btau}[#1]}
}
\NewDocumentCommand{\dplusr}{o}{
	\IfNoValueTF{#1}
	{\op{\mathcal D}{+}{\taur}}
	{\op{\mathcal D}{+}{\taur}[#1]}
}
\NewDocumentCommand{\dplusmell}{o}{
	\IfNoValueTF{#1}
	{\op{\mathcal D}{+}{\taumell}}
	{\op{\mathcal D}{+}{\taumell}[#1]}
}
\NewDocumentCommand{\Dplusmell}{o}{
	\IfNoValueTF{#1}
	{\op{\bs{\mathcal D}}{+}{\btaum_\ell}}
	{\op{\bs{\mathcal D}}{+}{\btaum_\ell}[#1]}
}
\NewDocumentCommand{\Dplusrell}{o}{
	\IfNoValueTF{#1}
	{\op{\bs{\mathcal D}}{+}{\btaur_\ell}}
	{\op{\bs{\mathcal D}}{+}{\btaur_\ell}[#1]}
}
\NewDocumentCommand{\Dplusrellmo}{o}{
	\IfNoValueTF{#1}
	{\op{\bs{\mathcal D}}{+}{\btaur_{\ell-1}}}
	{\op{\bs{\mathcal D}}{+}{\btaur_{\ell-1}}[#1]}
}
\NewDocumentCommand{\dplusmi}{o}{
	\IfNoValueTF{#1}
	{\op{\mathcal D}{+}{\taum_i}}
	{\op{\mathcal D}{+}{\taum_i}[#1]}
}
\NewDocumentCommand{\dplusmj}{o}{
	\IfNoValueTF{#1}
	{\op{\mathcal D}{+}{\taum_j}}
	{\op{\mathcal D}{+}{\taum_j}[#1]}
}
\NewDocumentCommand{\Dplusr}{o}{
	\IfNoValueTF{#1}
	{\op{\bs{\mathcal D}}{+}{\btaur}}
	{\op{\bs{\mathcal D}}{+}{\btaur}[#1]}
}
\NewDocumentCommand{\dplusw}{o}{
	\IfNoValueTF{#1}
	{\op{\mathcal D}{+}{\tauw}}
	{\op{\mathcal D}{+}{\tauw}[#1]}
}

\newcommand{\triplesup}[3]{%
  \mkern3mu\mathord{\vbox{\offinterlineskip\ialign{%
    \hfil##\hfil\cr
	$\scriptscriptstyle \vspace{1pt} \mathrm{#3}$\cr
	$\scriptscriptstyle \vspace{1pt} \mathrm{#2}$\cr
    $\scriptscriptstyle \mathrm{#1}$\cr
    \noalign{\kern1ex}
}}}}

\newcommand{\doublesup}[2]{%
  \mkern3mu\mathord{\vbox{\offinterlineskip\ialign{%
    \hfil##\hfil\cr
	$\scriptscriptstyle \vspace{1pt} \mathrm{#2}$\cr
    $\scriptscriptstyle \mathrm{#1}$\cr
    \noalign{\kern1ex}
}}}}

\newcommand{\bbh}[2]{\hat{\breve{\bar{#1}}}_{#2}\hspace{-5pt}\triplesup{r}{m}{m}}
\newcommand{\bbrm}[2]{\breve{\bar{#1}}_{#2}\hspace{-4pt}\doublesup{r}{m}}
\newcommand{\htrm}[2]{\tilde{\hat{#1}}_{#2}\hspace{-4pt}\doublesup{r}{m}}
\newcommand{\bhmm}[2]{\hat{\breve{#1}}_{#2}\hspace{-5pt}\doublesup{m}{m}}
\newcommand{\bhrr}[2]{\hat{\breve{#1}}_{#2}\hspace{-5pt}\doublesup{r}{r}}
\newcommand{\thmm}[2]{\hat{\tilde{#1}}_{#2}\hspace{-5pt}\doublesup{r}{r}}
\newcommand{\bbmm}[2]{\breve{\bar{#1}}_{#2}\hspace{-5pt}\doublesup{r}{r}}

\newcommand{\eb}{E^\mathrm{B}}
\newcommand{\edot}{\dot e}
\newcommand{\eell}{e_\ell}
\newcommand{\Eell}{\bm e_\ell}
\newcommand{\eg}{\hat E^\mathrm{B}}
\newcommand{\einst}{e^\text{inst}}
\newcommand{\Einst}{\bm e^\text{inst}}
\newcommand{\einstell}{\einst_\ell}
\newcommand{\Einstell}{\Einst_\ell}
\newcommand{\ep}{E^\mathrm{P}}
\newcommand{\epsilonbreve}{\widebreve{\epsilon}}
\newcommand{\epsilonhat}{\hat{\epsilon}}
\newcommand{\epsilontilde}{\widetilde{\epsilon}}
\newcommand{\etatheta}{\eta_{\bm \theta}}
\newcommand{\etaW}{\eta^W}
\newcommand{\EtaW}{\bm \eta^W}
\newcommand{\etaWell}{\etaW_\ell}
\newcommand{\EtaWell}{\EtaW_\ell}
\newcommand{\hatm}[1]{\widehat{#1}^{\text{m}}}
\newcommand{\hatr}[1]{\widehat{#1}^{\text{r}}}

\newcommand{\id}{\mathds{1}}

\NewDocumentCommand{\iminus}{o}{
	\IfNoValueTF{#1}
	{\op{\mathcal I}{-}{\tau}}
	{\op{\mathcal I}{-}{\tau}[#1]}
}
\newcommand{\iminushat}{\op{\widehat{\mathcal I}}{-}{\tau}}
\NewDocumentCommand{\iminusm}{o}{
	\IfNoValueTF{#1}
	{\op{\mathcal I}{-}{\taum}}
	{\op{\mathcal I}{-}{\taum}[#1]}
}
\NewDocumentCommand{\iminusmell}{o}{
	\IfNoValueTF{#1}
	{\op{\mathcal I}{-}{\taumell}}
	{\op{\mathcal I}{-}{\taumell}[#1]}
}
\NewDocumentCommand{\iminusmi}{o}{
	\IfNoValueTF{#1}
	{\op{\mathcal I}{-}{\taum_i}}
	{\op{\mathcal I}{-}{\taum_i}[#1]}
}
\NewDocumentCommand{\iminusrell}{o}{
	\IfNoValueTF{#1}
	{\op{\mathcal I}{-}{\taurell}}
	{\op{\mathcal I}{-}{\taurell}[#1]}
}
\NewDocumentCommand{\Iminusrell}{o}{
	\IfNoValueTF{#1}
	{\op{\bs{\mathcal I}}{-}{\btaur_\ell}}
	{\op{\bs{\mathcal I}}{-}{\btaur_\ell}[#1]}
}
\NewDocumentCommand{\iminusri}{o}{
	\IfNoValueTF{#1}
	{\op{\mathcal I}{-}{\taur_i}}
	{\op{\mathcal I}{-}{\taur_i}[#1]}
}
\NewDocumentCommand{\iminusrj}{o}{
	\IfNoValueTF{#1}
	{\op{\mathcal I}{-}{\taur_j}}
	{\op{\mathcal I}{-}{\taur_j}[#1]}
}
\NewDocumentCommand{\Iminusm}{o}{
	\IfNoValueTF{#1}
	{\op{\bs{\mathcal I}}{-}{\btaum}}
	{\op{\bs{\mathcal I}}{-}{\btaum}[#1]}
}
\NewDocumentCommand{\Iminusmell}{o}{
	\IfNoValueTF{#1}
	{\op{\bs{\mathcal I}}{-}{\btaumell}}
	{\op{\bs{\mathcal I}}{-}{\btaumell}[#1]}
}
\NewDocumentCommand{\iminusr}{o}{
	\IfNoValueTF{#1}
	{\op{\mathcal I}{-}{\taur}}
	{\op{\mathcal I}{-}{\taur}[#1]}
}
\newcommand{\iminusrhat}{\op{\widehat{\mathcal I}}{-}{\taur}}
\NewDocumentCommand{\Iminus}{o}{
	\IfNoValueTF{#1}
	{\op{\bs{\mathcal I}}{-}{\btau}}
	{\op{\bs{\mathcal I}}{-}{\btau}[#1]}
}
\NewDocumentCommand{\Iminusr}{o}{
	\IfNoValueTF{#1}
	{\op{\bs{\mathcal I}}{-}{\btaur}}
	{\op{\bs{\mathcal I}}{-}{\btaur}[#1]}
}

\newcommand{\intC}{\mathcal C}
\newcommand{\intE}{\mathcal E}

\NewDocumentCommand{\iplus}{o}{
	\IfNoValueTF{#1}
	{\op{\mathcal I}{+}{\tau}}
	{\op{\mathcal I}{+}{\tau}[#1]}
}
\NewDocumentCommand{\iplusm}{o}{
	\IfNoValueTF{#1}
	{\op{\mathcal I}{+}{\taum}}
	{\op{\mathcal I}{+}{\taum}[#1]}
}
\newcommand{\iplusmhat}{\op{\widehat{\mathcal I}}{+}{\taum}}
\NewDocumentCommand{\Iplusm}{o}{
	\IfNoValueTF{#1}
	{\op{\bs{\mathcal I}}{+}{\btaum}}
	{\op{\bs{\mathcal I}}{+}{\btaum}[#1]}
}
\NewDocumentCommand{\iplusmell}{o}{
	\IfNoValueTF{#1}
	{\op{\mathcal I}{+}{\taumell}}
	{\op{\mathcal I}{+}{\taumell}[#1]}
}
\NewDocumentCommand{\Iplusmell}{o}{
	\IfNoValueTF{#1}
	{\op{\bs{\mathcal I}}{+}{\btaum_\ell}}
	{\op{\bs{\mathcal I}}{+}{\btaum_\ell}[#1]}
}
\NewDocumentCommand{\IplusmL}{o}{
	\IfNoValueTF{#1}
	{\op{\bs{\mathcal I}}{+}{\btaum_L}}
	{\op{\bs{\mathcal I}}{+}{\btaum_L}[#1]}
}
\NewDocumentCommand{\iplusmi}{o}{
	\IfNoValueTF{#1}
	{\op{\mathcal I}{+}{\taum_i}}
	{\op{\mathcal I}{+}{\taum_i}[#1]}
}
\NewDocumentCommand{\iplusmj}{o}{
	\IfNoValueTF{#1}
	{\op{\mathcal I}{+}{\taum_j}}
	{\op{\mathcal I}{+}{\taum_j}[#1]}
}
\NewDocumentCommand{\iplusr}{o}{
	\IfNoValueTF{#1}
	{\op{\mathcal I}{+}{\taur}}
	{\op{\mathcal I}{+}{\taur}[#1]}
}
\newcommand{\iplushat}{\op{\widehat{\mathcal I}}{+}{\tau}}
\NewDocumentCommand{\Iplusr}{o}{
	\IfNoValueTF{#1}
	{\op{\bs{\mathcal I}}{+}{\btaur}}
	{\op{\bs{\mathcal I}}{+}{\btaur}[#1]}
}

\newcommand{\mustbe}{\overset!=}
\newcommand{\norm}[1]{\left\lVert #1 \right\rVert}

\NewDocumentCommand{\op}{mmmo}{
	\IfNoValueTF{#4}
	{{#1}^{#2}_{#3}}
	{{#1}^{#2}_{#3} \left\{ #4 \right\}}
}
\newcommand{\opd}{\mathcal{D}}
\newcommand{\opi}{\mathcal{I}}
\newcommand{\rell}{r_\ell}
\newcommand{\Rell}{\bm{r}_\ell}
\newcommand{\relu}{\text{ReLU}}

\usepackage{scalerel}[2014/03/10]
\usepackage{stackengine}
\newcommand\reallywidehat[1]{\ThisStyle{%
  \setbox0=\hbox{$\SavedStyle#1$}%
  \stackengine{-1.0\ht0+.5pt}{$\SavedStyle#1$}{%
    \stretchto{\scaleto{\SavedStyle\mkern.15mu\char'136}{2.6\wd0}}{1.4\ht0}%
  }{O}{c}{F}{T}{S}%
}}

\newcommand{\rin}{r_\text{in}}
\newcommand{\roundpar}[1]{\left( #1 \right)}
\newcommand{\rpre}{r_\text{pre}}
\newcommand{\rT}{r^\TT}
\newcommand{\taudot}{\dot\tau}
\newcommand{\taum}{\tau^\mathrm{m}}
\newcommand{\Taum}{\bm\tau^\mathrm{m}}
\newcommand{\taumell}{\tau^\mathrm{m}_\ell}
\newcommand{\Taumell}{\bm\tau^\mathrm{m}_\ell}
\newcommand{\taur}{\tau^\mathrm{r}}
\newcommand{\Taur}{\bm\tau^\mathrm{r}}
\newcommand{\taurell}{\tau^\mathrm{r}_\ell}
\newcommand{\Taurell}{\bm\tau^\mathrm{r}_\ell}
\newcommand{\tauw}{\tau^w}
\newcommand{\tauW}{\tau^W}
\newcommand{\TauW}{\bm\tau^W}
\newcommand{\tauWell}{\tauW_\ell}
\newcommand{\TauWell}{\TauW_\ell}
\newcommand{\tb}[1]{\textbf{#1}}
\newcommand{\thetadot}{\dot \theta}
\newcommand{\Thetadot}{\dot{\bm \theta}}
\newcommand{\tildem}[1]{\widetilde{#1}^{\text{m}}}
\newcommand{\tilder}[1]{\widetilde{#1}^{\text{r}}}
\newcommand{\TT}{\mathrm{T}}
\newcommand{\ubrevem}{\brevem{u}}
\newcommand{\ubrever}{\brever{u}}
\newcommand{\udot}{{\dot u}}
\newcommand{\Udot}{\dot{\bm u}}
\newcommand{\uell}{u_\ell}
\newcommand{\Uell}{\bm u_\ell}
\newcommand{\Uelldot}{\dot{\bm u}_\ell}
\newcommand{\vell}{v_\ell}
\newcommand{\Vell}{\bm v_\ell}
\newcommand{\Wdot}{\dot W}
\newcommand{\Well}{\bm W_\ell}
\newcommand{\WER}{W_\mathrm{ER}}

\makeatletter
\def\widebreve{\mathpalette\wide@breve}
\def\wide@breve#1#2{\sbox\z@{$#1#2$}%
	\mathop{\vbox{\m@th\ialign{##\crcr
				\kern0.08em\brevefill#1{0.8\wd\z@}\crcr\noalign{\nointerlineskip}%
				$\hss#1#2\hss$\crcr}}}\limits}
\def\brevefill#1#2{$\m@th\sbox\tw@{$#1($}%
	\hss\resizebox{#2}{\wd\tw@}{\rotatebox[origin=c]{90}{\upshape(}}\hss$}
\makeatletter

\newcommand{\WRE}{W_\mathrm{RE}}
\newcommand{\WITH}{\mathrm{with}}
\newcommand{\WT}{W^\TT}

\renewcommand{\odot}{\circ}

\title{A Variational Latent Equilibrium for Learning in Neuronal Circuits}

\author[1]{Simon Brandt}
\author[1]{Paul Haider}
\author[1]{Walter Senn}
\author[1,+,*]{Federico Benitez}
\author[1,+]{Mihai A. Petrovici}

\affil[1]{Department of Physiology, University of Bern, Bühlplatz 5, 3012, Bern, Switzerland}

\affil[+]{Joint senior authorship}

\affil[*]{federico.benitez@unibe.ch}

\begin{document}

\maketitle

\begin{abstract}
Brains remain unrivaled in their ability to recognize and generate complex spatiotemporal patterns.
While AI is able to reproduce some of these capabilities, deep learning algorithms remain largely at odds with our current understanding of brain circuitry and dynamics.
This is prominently the case for backpropagation through time (BPTT), the go-to algorithm for learning complex temporal dependencies.
In this work we propose a general formalism to approximate BPTT in a controlled, biologically plausible manner. 
Our approach builds on, unifies and extends several previous approaches to local, time-continuous, phase-free spatiotemporal credit assignment based on principles of energy conservation and extremal action.
Our starting point is a prospective energy function of neuronal states, from which we calculate real-time error dynamics for time-continuous neuronal networks.
With this, as a first contribution we provide a simple and straightforward derivation of the adjoint method dynamical equations, the time-continuous equivalent to BPTT.
Importantly, additional approximations are necessary to avoid unphysical temporal non-localities. 
One such approximation was recently proposed in Generalized Latent Equilibrium (GLE), but enforcing temporal locality of the backpropagating dynamics induces an error in the gain that becomes problematic when processing complex signals.
Thus, our second contribution is to design a learning rule for the backward weights to find a fully local (in space and time) set of equations for neuron and synapse dynamics that corrects for this gain difference.
With this inclusion of backward weight learning, our theory provides a rigorous framework for spatiotemporal deep learning in the brain, while simultaneously advancing a blueprint for physical circuits capable of carrying out these computations.
\end{abstract}

\section{Introduction}

The brain excels at identifying and generating complex patterns across space and time.
When learning to perform such kind of task, synaptic connections between neurons, and thus the brain's behavior, are changed to increase performance.
Performance can be described by a cost function that measures how well the behavior aligns with a target behavior: to achieve optimal performance, system parameters need to change such that the cost is minimized.
Notably, this optimization has to take place under the physical and biological constraints of the neuronal system -- a central one being locality.
Minimizing the cost while forcing the dynamics of the system to obey pre-defined physical and biological constraints constitutes a constrained optimization problem.
A general method to solve such problems applies calculus of variations, taking inspiration from physics and from the theory of optimal control.
It is important to emphasize that in biology, such an optimal solution needs to be learned by a network of physical neurons.

The classical treatment of such problems in \ml and computational neuroscience are respectively  \bptt \cite{pineda1987generalization, werbos1990backpropagation}, and \rtrl~\cite{williams1989learning}.
These methods provide the exact gradients for learning spatio-temporal tasks, but have major shortcomings as models of biological processes. \bptt requires perfect knowledge of the future states of the network, which clearly violates causality.
In contrast, while \rtrl conserves causality, it requires the in-memory storage of all past influences of every synaptic weight on all neuronal activities, which also strains biological plausibility (as well as practical implementations).

Recently, \gle~\cite{ellenberger2025backpropagation} proposed a biologically plausible algorithm, local in space and time, to implement spatio-temporal learning in physical networks of neurons.
This framework extends and generalizes \le~\cite{haider2021latent} to enable networks to perform temporal processing tasks.
These frameworks leverage the temporal processing capabilities of biological neurons in two ways. 
As a first ingredient, neuronal membrane integration performs low-pass filtering of the input, with the associated memory effects and lagged signals.
Additionally, it is well-established though less well-known that neurons can generally react prospectively to their input, so that their output is based on the expected future input. 
For a more detailed discussion of temporal processing at the level of individual neurons we refer the reader to~\cite{brandt2024prospective}.
These two temporal operations are implemented locally at the level of individual neurons, leading to biologically plausible neuron dynamics and learning through synaptic updates.
\gle can be shown to approximate the \am, which is the standard method to solve constrained optimization problems and the equivalent of \bptt in continuous time settings.
While \le deals with purely spatial problems, and is related to methods such as equiprop~\cite{scellier2017equilibrium}, and the neuronal least action principle~\cite{senn2024neuronal}, \gle solves spatiotemporal problems and is related to \bptt and \rtrl.

In this work, we extend and improve these previous results.
First, we go back to a first-principles approach in terms of variational calculus, which enables a unified description of many of these related methods.
We give an explicit derivation of neuronal dynamics for systems of neurons equipped with these temporal processing capabilities.
This derivation recovers the results of the \am, while being much simpler and conceptually more transparent than standard derivations.
\am is not an established method for approaching network dynamics in computational neuroscience, and our approach aims to overcome this gap by improving accessibility. 
Furthermore, our approach anchors prospective coding -- a ubiquitous phenomenon in biology, but that has received little attention in computational studies -- in the calculus of variations. 
In order to recover biological plausibility, we show which explicit approximations within our formalism give rise to \gle dynamics.
In its original version, \gle is not as strongly grounded in the calculus of variations, but rather derives network dynamics from a stationarity postulate.
As a second contribution, we cash out this conceptual transparency to propose an explicit learning rule for backward synapses, that provides a better approximation of the \am compared to \gle, while remaining biologically plausible (synapse-local in space and time).
Our new approach, which we dub \vle, thus goes beyond a more rigorous reframing of \gle, but represents a genuine extension. 
We provide some promising first applications in learning to reproduce complex temporal behavior.

\section{Results}
We consider a general setup for a neuronal network,  capable of representing systems of physical neurons such as biological and neuromorphic systems. We denote the input signal by $\bm x(t)$, and the output of the network by $\bm y(t)$. In a feedforward network this output will be generated by the last layer of the network. In the specific case of supervised learning, the target output is denoted by the vector $\bm y^*(t)$, and the instantaneous cost function by $C\big(\bm y(t), \bm y^*(t) \big)$. Networks depend on parameters denoted by the vector $\bm \theta$, which include, for example, synaptic weights $\bm W$ and time constants $\bm \tau$.

In the most general case of a temporal processing task, we want to minimize the integrated cost $\intC = \int_{t_1}^{t_2} \dd t \, C(t)$ with respect to the parameters $\bm \theta$, by means of some implementation of gradient descent:
\begin{equation}
	\Thetadot \sim - \bm \nabla_{\bm \theta} \intC\,.
\end{equation}
As the dynamics of the network are constrained by its physical properties, this constitutes an example of a \emph{constrained optimization} problem \cite{kelley1960gradient,todorov2006optimal,chachuat2007nonlinear}.

A general solution to this problem is provided by the powerful \am~\cite{kelley1962method}, but its usual derivation is relatively involved, and does not take into account the specific characteristics of bio-inspired neuronal dynamics.
Here we reach the same outcome as the \am while using a simple derivation that is better suited to the needs of physical networks, whether biological or neuromorphic.
To achieve this, we harness the dual power of (i) an energy-based approach inspired by the many results in physics that use the minimization of a function to encode physical dynamics, and (ii) a specific functional form for the constraints, inspired by the phenomenon of prospectivity in biological neuronal networks \cite{brandt2024prospective}.
Our derivation encompasses and extends previous results such as \le~\cite{haider2021latent}, and \gle~\cite{ellenberger2025backpropagation}, while being easily adaptable to also derive the \nla principle~\cite{senn2024neuronal}, all of which can be shown to be related to our overarching formalism.

 In the following, we use scalars rather than vectors to highlight locality, a key factor in bio-plausibility.
 Following assumption (i), we postulate a network energy $E$ that is a sum over squared neuron-local errors $e_i$ and the global cost $C$ multiplied by a small nudging parameter $\beta$:
\begin{equation}\label{total_energy}
	E(t) = \frac12 \sum_i e_i^2(t) + \beta C(t)\,.
\end{equation}
The specific form of these errors ultimately determines the structure and dynamics of the system. As we show in the Supplement, for $\beta \to 0$ the minimization of this energy with respect to the parameters $\bm \theta$ is equivalent to the minimization of the cost $C$.

For networks of leaky integrator neurons, we define the local mismatch error as
\begin{equation}
	e_i = \ubrevem_i - \sum_j W_{ij} \varphi_j(\ubrever_j)\,.
	\label{eqn:edef}
\end{equation}

$W_{ij}$ are the synaptic input weights and the output rate of a neuron is defined as $r_i = \varphi_i(\ubrever_i)$, with activation $\varphi_i$ a function of the neuron's membrane voltage $u_i$. Throughout this manuscript, we use  a series of shorthands for temporal operators, as described in \cref{tbl:operators}.

\begin{table}[h]
    \small
	\scriptsize
	\bgroup
	\setlength{\tabcolsep}{2pt}
	\begin{tabular}{l|ll}
		 & exponential filtering & temporal differentiation \\\hline
		\makecell[tl]{forward-\\facing} & \makecell[tl]{discounted future\\$\tildem{u} = \frac 1 \taum \int_{t}^\infty u(t') \exp \left( \frac{t-t'}{\taum} \right) \dd t'$} & \makecell[tl]{look-ahead\\$\brever{u} = \left( 1 + \taur \frac{\dd}{\dd t} \right) u(t)$} \\
		\makecell[tl]{backward-\\facing} & \makecell[tl]{low-pass\\$\barm{u} = \frac 1 \taum \int_{-\infty}^t u(t') \exp \left( \frac{t'-t}{\taum} \right) \dd t'$} & \makecell[tl]{look-back\\$\hatr{u} = \left( 1 - \taur \frac{\dd}{\dd t} \right) u (t)$}
	\end{tabular}
	\egroup
	\caption{The temporal operators can be separated into taking the future (forward facing) or past (backward facing) into account }
	\label{tbl:operators}
\end{table}

These operators represent retrospective (past-facing) as well as prospective (future-facing) behaviors, both observed in biological neurons.
Briefly stated, the low-pass filter operator stems from membrane integration, and the look-ahead operator represents the ability of cortical neurons to be sensitive not only to the current input, but to how this input is changing, and to react prospectively to that information.
$\taum$ and $\taur$ represent the timescales for membrane integration and output prospectivity respectively.
It is easy to check that these operators are pairwise inverses of each other:
\begin{equation}
	\thmm{u}{\,\,\,} = \bbmm{u}{\,\,\,} = u.
	\label{eq:operators_cancel}
\end{equation}

For learning, we derive local plasticity rules by gradient descent. Locality of plasticity follows by virtue of the locality of the individual neuronal energies $E_i = \frac 1 2 e_i^2$ defined above. The plasticity rules read:
\begin{equation}\label{plasticity}
	\thetadot_i \sim -\frac{\partial E}{\partial \theta_i} =  -e_i \frac{\partial e_i}{\partial \theta_i}
\end{equation}
as well as a possible additional nudging term proportional to $\beta$ -- usually only present in output neurons during training. The index of $\theta_i$ indicates its association to neuron $i$ and thereby its unique appearance in each $E_i$.

In particular, for synaptic weights, we get dynamics
\begin{equation}\label{Wlearning}
	\dot W_{ij} = e_i  r_j \, .
\end{equation}

We now derive error propagation from the energy as well; to account for the integrated cost $\intC$, we need to consider the integrated energy\footnote{
	As energy-based approaches such as ours are often inspired by physics, they have sometimes adopted further terminology. For example, the energy has also been called a Lagrangian, and its integral an action in, e.g., \cite{senn2024neuronal}. The term ``Lagrangian'' also appears in the context of optimization, but rather as a link to the Lagrange multiplier method (e.g., in \cite{boyd2004convex, luenberger1984linear}). Here, we eschew this terminology in order to avoid stronger implications from theoretical physics that do not apply here.
} $\intE = \int_{t_1}^{t_2} \dd t \, E(t)$.

Following standard methodology, we require neuronal dynamics to keep the integrated energy stationary:
\begin{equation}
	\delta_u \intE = 0.
\end{equation}
This variational approach can be viewed as a form of equilibrium in the trajectory of lowest integrated energy rather than just the lowest instantaneous energy.
Furthermore, as we show in more detail below, the outcome of the method links directly to (G)LE.
Because of these reasons, we call this approach Variational Latent Equilibrium, \vle. 

Finding the extremum of $\intE$ is equivalent to having neuronal dynamics obey the corresponding Euler-Lagrange equations, which yield the error dynamics (see Supplement for details)
\begin{align}
	e_i  = \tilde\varepsilon_i^\mathrm{m} \; \textrm{with}\; \;  \varepsilon_i = {\reallywidehat{\varphi'_i \sum_k W_{ki} e_k}}^\mathrm{r},
	\label{eqn:eproptle}
\end{align}
which is exactly the result of the \am for this type of system~\cite{ellenberger2025backpropagation}. Note that this is derived in a much simplified way compared to standard treatments in the context of optimal control theory, thanks to the specific form taken by dynamical variables when modeling prospective behavior. It is remarkable that one can reproduce these results from first principles at a small fraction of the usual analytical cost, given that the range of application of our dynamics is still ample within computational neuroscience and neuromorphic computing.  

As in standard \am, this expression for error propagation violates causality, because the discounted future error depends on future states. Solutions to this issue usually require biologically implausible treatments. 
Additionally, the expression also violates spatial locality, because backward errors $e_j$ need to propagate through forward synapses $W_{ji}$, an expression of the classical weight transport problem. Below we show how to deal with both these issues. 

To explicitly see how errors couple into neuronal dynamics, we can simply rearrange \cref{eqn:edef} to obtain
\begin{equation}
	\taum \udot_i = - u_i + \sum_j W_{ij} \varphi_j(\ubrever_j) + e_i
	\label{eqn:udot}
\end{equation}
with ``forward" propagation of signals $\varphi_j(\ubrever_j)$ and ``backward" propagation of the errors $e_k$ in \cref{eqn:eproptle}.

A natural interpretation of \cref{eqn:udot} corresponds to a three-compartment neuron, with a somatic compartment storing $u_i$, an input compartment storing $\sum_j W_{ij} \varphi_j(\ubrever_j)$ and an error compartment storing $e_i$; in cortical neurons, a likely (but not strictly necessary) assignment of such compartments would be for the input compartment to be identified as basal and the error compartment as apical. Note that in the \am, neuronal activity is not affected by errors; in our networks, this is reproduced in the limit of $\beta \to 0$ (see Supplement).

At this point we can easily consider the special case $\taum=\taur$; then, as discussed above \cref{eq:operators_cancel}, the discounted future and look-back operators cancel each other out and \vle reduces to \le \cite{haider2021latent}: $e_i = \varphi'_i(\ubrevem_i) \sum_j W_{ji} e_j$. This is equivalent to requiring that neuronal dynamics conserve the instantaneous energy $\partial_u E(t) = 0$ (which is indeed the starting point of \le), which yields the instantaneous errors necessary to reduce the instantaneous cost.\footnote{
		These instantaneous equations are also related to those derived in the \nla framework \cite{senn2024neuronal}, but with an important difference: in \nla, instantaneity is achieved by having each neuron use its low-pass filter to undo its presynaptic partners' prospective filters.
		Thus, for each neuron, all its postsynaptic partners need to share their $\taum$ and all its presynaptic partners their $\taur$ (i.e., given a network graph, all neurons within a partition -- for example, within a layer -- need to share time constants).
		Furthermore, since neurons receiving an external input necessarily carry out a low-pass filter that is never compensated for, the network cannot react in the instantaneous manner required by exact error backpropagation.
	}
	Importantly, in \le there is no issue with causality, as the dynamics only depend on the instantaneous state and velocity of the neurons and error signals. 

Setting aside this special case, to achieve temporal locality in the error stream we approximate the discounted future operator by the look-ahead (see \Cref{tbl:operators}); additionally, we can approximate the look-back by the low-pass filter to make error dynamics even more neuron-like, which has the additional advantage of partially compensating for the approximation made to the forward signal \cite{ellenberger2025backpropagation}. With these approximations, we get
\begin{align}
	e_i \simeq  \breve{\epsilon}_i^\mathrm{m} \;  \textrm{with}\; \;  \epsilon_i  \simeq  {\overline{\varphi'_i \sum_k W_{ki} e_k}}^{\;\mathrm{r}},
	\label{eqn:epropgle}
\end{align}
and recover the \gle equations~\cite{ellenberger2025backpropagation}.
The best way to understand this approximation is through Fourier analysis which yields the Fourier transforms and the corresponding gains for each operator, as shown in \cref{tbl:operators-omega}.
The approximations towards \gle conserve the phase shift of all Fourier components in the filtered signal while maintaining causality, at the cost of distorting their amplitudes by a gain factor $\sqrt{1 + \roundpar{\omega \tau}^2}$ for each of the two operator switches.

\begin{table}[h]
    \center
    \small
	\footnotesize
	\bgroup
	\setlength{\tabcolsep}{2pt}
	\begin{tabular}{l|cccc}

		 & $\hat{x}^\mathrm{m}(t)$ & $\breve{x}^\mathrm{m}(t)$ & $\bar{x}^\mathrm{m}(t)$ & $\tilde{x}^\mathrm{m}(t)$ \\
		\hline
		$\mathcal{H}(\omega)$ & $1 - i\omega\tau_\mathrm{m}$ & $1 + i\omega\tau_\mathrm{m}$ & $\dfrac{1}{1 + i\omega\tau_\mathrm{m}}$ & $\dfrac{1}{1 - i\omega\tau_\mathrm{m}}$ \\
		$\mathcal{G}(\omega)$ & $\sqrt{1 + (\omega\tau_\mathrm{m})^2}$ & $\sqrt{1 + (\omega\tau_\mathrm{m})^2}$ & $\dfrac{1}{\sqrt{1 + (\omega\tau_\mathrm{m})^2}}$ & $\dfrac{1}{\sqrt{1 + (\omega\tau_\mathrm{m})^2}}$
	\end{tabular}
	\egroup
	\caption{The temporal operators can directly be translated into Fourier space for each single frequency component $\omega$. The Fourier transform $\mathcal{H}(\omega)$ yields the gain $\mathcal{G}(\omega)$ for each operator.}
	\label{tbl:operators-omega}
\end{table}

\begin{figure}[t!]
    \centering
    \includegraphics[width=1\linewidth]{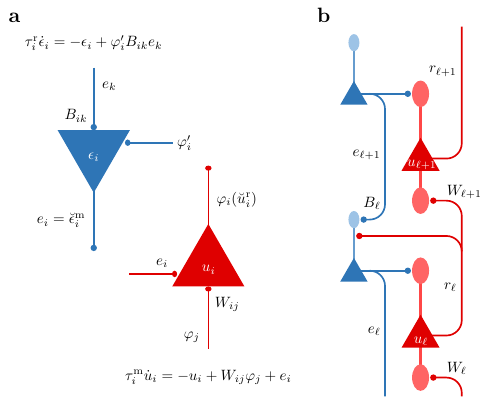}
	\caption{
		\tb{A symmetric microcircuit for spatially and temporally aligned backpropagation of errors.}
		\tb{(a)} Signal (bottom-up) and error (top-down) neurons, their incoming and outgoing firing rates and their dynamics.
        \tb{(b)} A chain of two neurons as the simplest network structure for studying error backpropagation.
    }
    \label{fig:circuit}
\end{figure}

To see how \vle can go beyond this approximation we first discuss the issue of spatial locality, the (in)famous weight transport problem. For this, we introduce backward weights $B_{ik}$  
 \begin{align}
    e_i & = \breve{\epsilon}_i^\mathrm{m} \; & \textrm{with}\; \;  \epsilon_i & = {\overline{\varphi'_i \sum_k B_{ik} e_k}}^{\;\mathrm{r}} \;.
	\label{eq:epropgle}
\end{align}
By relaxing the requirement of symmetric backward weights  $B = W^T$, the neurons required to implement the signal and error dynamics are symmetric in their temporal processing by low-pass filtering input and advancing their output prospectively, while having different weights in the forward and backward path and respecting spatial locality, as shown in \cref{fig:circuit}a and b.

As a first approximation, backward weights can be random and constant, so that the learning of the forward weights can compensate for the deviations in the error signal -- also known as Feedback Alignment (FA) \cite{lillicrap2016random}.
Unfortunately, this solution is of limited help for multi-layered networks, where weight learning cannot compensate for the accumulation of errors across several layers of forward and backward weights.
More robustly, the backward weights could learn to mirror their reciprocal forward synapses $W_{ji}$ through specific learning rules, e.g., \pal \cite{max2024learning} (but see also \cite{akrout2019deep} for rate-based models such as we discuss in this paper, and \cite{guerguiev2019spikebased} and the recent \cite{gierlich2026spikebased} for the spiking case).

\begin{figure*}
    \centering
    \includegraphics[width=1\linewidth]{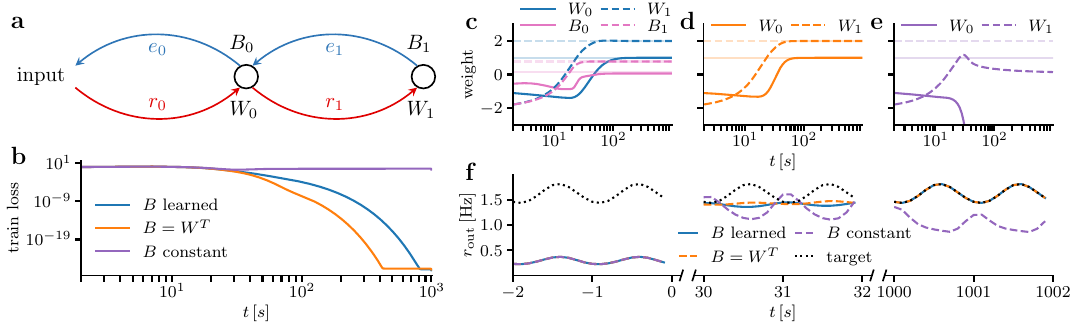}
	\caption{
		\tb{Learning a simple chain in a student-teacher setup.}
        \tb{(a)} Network structure, abstracted from \cref{fig:circuit}b for simplicity, with error and signal pathways indicated.
		\tb{(b)} The network converges to a minimal loss if the backward weights $\bm B$ are learned (blue) or fixed to $\bm{B}=\bm{W}^T$ (orange).
		For constant $\bm B$ (purple), the network does not converge.
		We plot a moving average of the loss to remove high-frequency oscillations.
		\tb{(c-e)} The student (opaque) and teacher (transparent) weights.
		Student $\bm W$ converge to the teacher weights if $\bm B$ is learned (c) or $\bm{B}=\bm{W}^T$ (d) and diverge for constant $\bm B$ (e).
		The learned backward weights $\bm B$ (c, pink) converge to the expected value given the time constant of the neuron and the frequency of the signal.
		\tb{(f)} compares the output of the three scenarios before learning, during learning and after learning to the target signal.
		}
    \label{fig:lagline}
\end{figure*}

\vle uses backward learning to improve on the approximations made by \gle while still respecting temporal as well as spatial locality. By learning backward weights that compensate for the gain offset in \gle, the network can better converge to the exact \am solution, and the dynamics can remain local. 

To see how this works, we start by defining instantaneous errors for \am and \vle, and write down how they propagate:
\begin{align}
	\text{\am}: e_i & = \tilde\varepsilon_i^\mathrm{m} \; & \textrm{with}\; \;  \varepsilon_i& = {\reallywidehat{\varphi'_i \sum_k W_{ki} e_k}}^\mathrm{r} \\
	\text{\vle}: e_i & = \breve{\epsilon}_i^\mathrm{m} \; & \textrm{with}\; \;  \epsilon_i & = {\overline{\varphi'_i \sum_k B_{ik} e_k}}^{\;\mathrm{r}}.
\end{align}
To equalize the two, we can learn a $\bm B$ that compensates for the difference between the look-ahead and the discounted future operators, so that $e_i^\text{\am} \mustbe e_i^\text{\vle}$. In fact we can consider the even stronger constraint:
\begin{equation}
	\sum_k W_{ki} \tilde\varepsilon_k^\mathrm{m} \mustbe \sum_k B_{ik} \breve{\epsilon}_k^\mathrm{m}.
\end{equation}

To achieve the above, we simply use local gradient descent to minimize
\begin{align}
	\eg_{i} &= \frac12 \roundpar{\sum_k W_{ki}  \bhrr{\epsilon}{k} - B_{ik} \bhmm{\epsilon}{k}}^2 \\
	 \dot B_{ij} &\sim - \frac{\partial \eg_{i}}{\partial B_{ij}} = \roundpar{\sum_k W_{ki}  \bhrr{\epsilon}{k} - B_{ik} \bhmm{\epsilon}{k}} \bhmm{\epsilon}{j}\,,
	\label{eqn:gal}
\end{align}
where both look-aheads and look-backs are causal and thus local in time (see Supplement for a detailed derivation).
We used $\varepsilon_j = \bhrr{\epsilon}{j}$ to compensate for the difference between the low-pass filter in \vle and the look-back operator in \am.
Since using $W_{ki}$ in the update of $B_{ij}$ requires weight transport, which is biologically not plausible, we point out that this expression must be combined with \pal to obtain a learning rule that is fully local in space and time (see Supplement for more details). In what follows we implement the \vle algorithm in some practical applications.

\tb{Lagline}

To demonstrate how learning of the backward weights $\bm B$ influences the training of networks, we use a simple chain of two neurons in a student-teacher setup as depicted in \cref{fig:lagline}a.
In this task, a student network has to adapt its weights to approximate the output of a teacher network with the same network topology.
Because of the simplicity of the network and the choice of nonlinearities, the task has a unique solution -- learning the teacher weights.
The weights of the student are initialized as $W_{ij}=-W_{ij}^*$, with $W_{01}^*$ and $W_{12}^*$ the non-zero teacher weights.
The results are shown in \cref{fig:lagline}, where we validate three scenarios: learning $\bm B$ with \cref{eqn:gal} (blue), using $\bm B=\bm{W}^T$ (orange) and $\bm B=\mathrm{const}$ (purple).
While the former two converge towards a loss minimum, the latter corresponds to Feedback Alignment with wrongly initialized weights and does not converge (\cref{fig:lagline}b).
Consistently, the weights converge towards the teacher weights for learning $\bm B$ (\cref{fig:lagline}c) and using $\bm B=\bm{W}^T$ (\cref{fig:lagline}d), but not for $\bm B=\mathrm{const}$ (\cref{fig:lagline}e).
Because the input to the network is a simple sine, we can estimate the target for the learned backward weights, to which their value $\bm B$ indeed converges (\cref{fig:lagline}c).
Since we initialized the weights different to the teacher network, the output for each scenario is off the target before learning (\cref{fig:lagline}f).
During learning, the outputs approach the target while the weights are optimized.

After learning, the scenarios of learning $\bm B$ and using $\bm B=\bm{W}^T$ converge to the target while $\bm B=\mathrm{const}$ does not converge to a good solution.
While we claim that learning $\bm B$ is beneficial for the learning of the forward weights, in this simple task, $\bm B=\bm{W}^T$ converges to the teacher weights faster.
Our algorithm requires more time to converge, as every single synapse has to independently learn $\bm W$ and $\bm B$.
To demonstrate the strength of our method to effectively induce a synapse-specific learning rate through the learning of $\bm B$, we explore more complex tasks that involve larger networks of neurons.

\tb{Lagnet}

\begin{figure*}[tbp]
    \centering
	\includegraphics[width=1\linewidth]{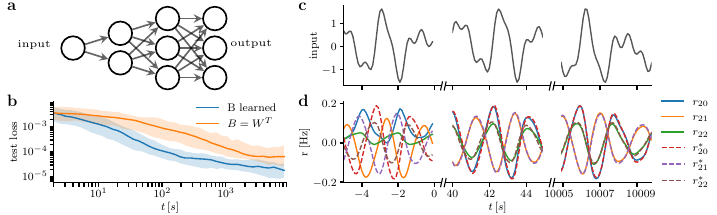}
    \caption{
		\tb{Learning in a network of neurons.}
		\tb{(a)} Forward network structure (error neurons omitted for readability).
		\tb{(b)} Network loss over time, evaluated on multiple test signals with randomly sampled frequencies and over multiple seeds.
        Orange: backward weights follow exact transposes of the forward weights.
        Blue: backward weights learned using our local update rule, \cref{eqn:gal}.
        Lines show median and the range 25$th$ to 75$th$ percentile of 60 seeds.
		\tb{(c)} Training input, consisting of multiple overlaid sines of different frequency, presented to the network at different times.
		\tb{(d)} Output of the student network learning $B$ (solid) compared to the teacher network (dashed) before training, during training and after training.
	}
    \label{fig:lagnet}
\end{figure*}

To investigate the performance of our model when solving more involved problems, we train a network of neurons with multiple outputs, as shown in \cref{fig:lagnet}a, again in a teacher-student setup.
Here, both the student and the teacher receive an input signal that is a combination of sines with different frequency and amplitude as shown for different times during training in \cref{fig:lagnet}c.
The network has two hidden layers with two and three neurons and three output neurons.
For such a network there is no longer a unique solution, and the learned weights can differ from the teacher weights.
Before learning, the output of the student differs from the output of the teacher (\cref{fig:lagnet}d).
During learning, the output of the student approaches the output of the teacher to finally converge after learning.

To test the performance of the model, we compare the scenario with learned $\bm B$ to using $\bm B=\bm{W}^T$.
Both scenarios are trained using only the signal shown in \cref{fig:lagnet}c and are tested on signals that consist of combinations of sines with randomly sampled frequencies within the frequency range of the training signal.
Initially, as compared to learning $\bm B$, the test loss using $\bm B=\bm{W}^T$ decreases faster, because of the higher learning rate at each synapse, to eventually stagnate at a relatively high test loss.
If we learn $\bm B$, the test loss decreases slower initially, but eventually reaches a lower minimum.
Thus, in this more complex task we can demonstrate that the synapse dependent learning rate obtained by learning $\bm B$ helps to converge to a better test performance.

\tb{Cartpole}

\begin{figure}[t!]
    \centering
    \includegraphics[width=1\linewidth]{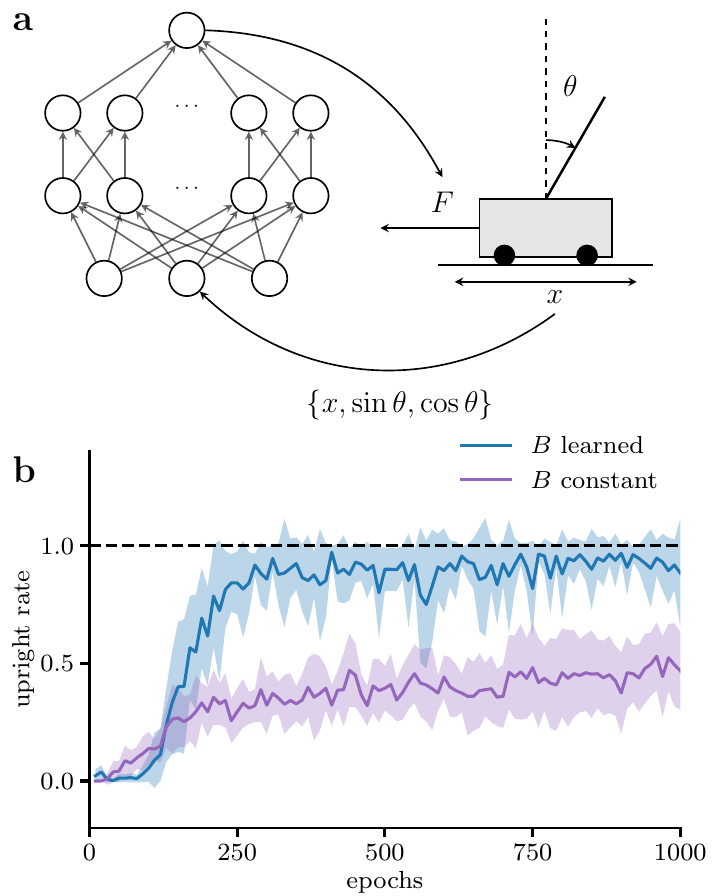}
    \caption{
		\tb{Cartpole task.}
        \tb{(a)} In this task, a pole on a cart has to be swung up and balanced to stay upright by applying a force $F$ on the cart.
        The network consists of three layers: an input layer getting only the positional signals $x$, $\sin(\theta)$ and $\cos(\theta)$, two hidden layers and an output layer with one neuron that yields the force to control the cartpole.
        \tb{(b)} When learning the backward weights, the network can achieve an almost perfect upright rate, defined as finishing the task with an upright pole.
        With constant backward weights, the performance is much worse and does not converge to high values.
	}
    \label{fig:cartpole}
\end{figure}

As a more complex application of \vle, we implemented a classical control task in the form of a $1D$ cartpole as described in \cite{barto1983neuronlike} and following the implementation by \cite{towers2025gymnasium}.
In this task, the network is trained to control the movement of a cart (see \Cref{fig:cartpole}a) by applying a force with the goal to balance the pole upright, while keeping the cart within pre-defined boundaries.
Solving this task usually requires to use position $x$ and angle ${\theta}$ as well as their temporal derivatives $\dot{x}$ and $\dot{\theta}$ as an input.
To enforce temporal credit assignment, we restrict the input of our network to position and angle only.
Excluding velocity information requires our network to generate latent variables that encode the relevant temporal information.
Because our method is currently only implemented in a supervised setting, we train a \vle network to mimic the output of an MPC controller \cite{rawlings2017model}.
Such controllers rely on partial observations and require to keep a memory of the past, as well as a prediction of the future.
To reproduce this behavior the network must then encode and process spatial and temporal information in a highly non-trivial manner.

In \Cref{fig:cartpole}b we compare the effectiveness of two different training protocols. 
For full \vle networks with $B$ learning, the network consistently learns to balance the pole across multiple runs.
For networks with fixed random backward weights, the network struggles to reach a high level of task fulfilment.
Notice that this latter case is equivalent to \gle with feedback alignment. 
Together, these results show the power of $B$ learning and the capabilities of \vle networks when considering complex temporal tasks.

\tb{\mnistId}

In order to test \vle on a larger, more challenging benchmark, we consider the \mnistId \cite{greydanus2024scaling} benchmark for temporal sequence classification. \mnistId is a temporal processing task as opposed to the more widely known MNIST, which is merely spatial. Each sample is a one-dimensional array of floating-point values, which is streamed as a temporal sequence into the network, to be classified among one of ten classes. This deceptively simple setup is hampered by two additional complications. First, only a quarter of the sample contains meaningful information, with the relevant part positioned randomly within a featureless padding carrier signal. Second, independent noise is added to the samples with multiple autocorrelation timescales, making the noise resistant to simple filters. 

\begin{figure}[tbp]
    \centering
    \includegraphics[width=1\linewidth]{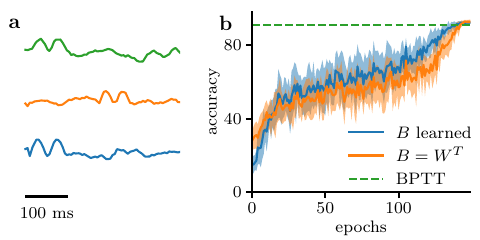}
    \caption{
		\tb{\mnistId.}
        \tb{(a)} The input time series encode one of the ten digits (in this case the digit $3$), but only a small part of the signal is relevant for classification.
        \tb{(b)} Validation accuracy over training epochs for the classification task using both $B$-learning (blue) and $B=W^T$ (orange), compared with BPTT (dashed green).
	}
    \label{fig:mnist1d}
\end{figure}

In \cref{fig:mnist1d}a we show a sample of the dataset. It is clear that online learning of \mnistId requires the network to access information about the past over timescales much longer than those of neuronal low-pass filtering (which are on the order of ten milliseconds). \cref{fig:mnist1d}b shows learning using \vle and backward weights set up as $B=W^T$ -- the equivalent to \gle. \gle is already close to the results stemming from \bptt, so the ultimate performance cannot be improved. However, we use this experiment to demonstrate that learning the backward weights can deal with such a challenging dataset and even shows slight improvements in the learning speed compared to using transpose weights.
This also constitutes a promising result for datasets where complex timescale interaction might make it challenging for \gle to find an optimal solution. This is the case in noisy systems, which we discuss next.

\tb{Noise injections}

\begin{figure*}[tbp]
    \centering
    \includegraphics[width=1\linewidth]{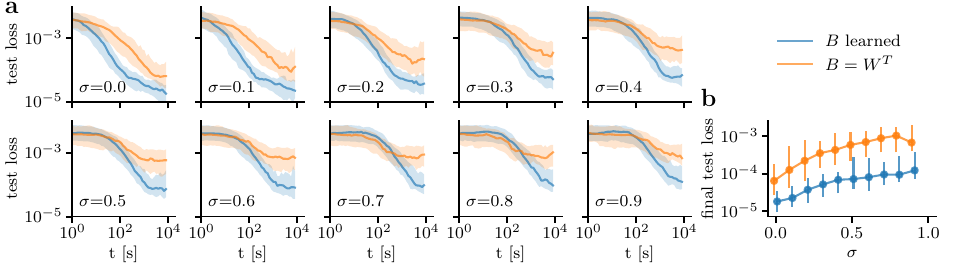}
    \caption{
		\tb{Lagnet with noise.}
        \tb{(a)} Test loss for a network with 5 layers under white noise of increasing variance for $B$ learned (blue) and $B=W^T$ (orange, equivalent to \gle).
        The lines show median and 25$th$ and 75$th$ percentile uncertainty of 60 seeds.
        \tb{(b)} The final test loss of the different noise conditions shown in (a), demonstrating a clear advantage for \vle.
	}
    \label{fig:noise}
\end{figure*}

To further stress the network, we constructed our own scenario as follows: we injected white noise into our lagnet experiments at increasing variance $\sigma$, and varied (specifically increased) the network’s size and depth (see Supplement).
Remember that \mnistId samples already include noise on a wide frequency band, but at a much lower amplitude than these experiments.
As can be seen in \cref{fig:noise}, at increasing noise levels \gle is faced with increasing amplitudes at higher frequencies, and its performance drops. In contrast, the $\mathbf B$-learning rule of \vle can adapt by decreasing synaptic weights and reducing high-frequency gains, allowing the network to maintain its performance.

We have scaled these simulations to up to $5$ layers and $256$ neurons per layer, and have observed how \vle consistently and increasingly outperforms \gle as the noise spectrum becomes broader, in particular in deeper networks where multiple layers of prospective neurons cumulatively increase the gains of high-frequency components. In the most extreme cases, the observed losses after learning for \vle are roughly one order of magnitude lower than for \gle, while remaining approximately on par with those from \bptt.

This robustness with respect to noise constitutes an important improvement over previous methods, and evinces \vle's advantages over \gle. In \gle an approximately quasi-static condition is required for good performance. \vle lifts this requirement because it actively counteracts the gain distortions that happen when violating this condition. In the simulations described above, we show clearly how a \vle network is able to deal with broad-spectrum noise, including especially high frequencies, without significant losses in performance. This robustness adds an essential ingredient to the model’s plausibility in analog systems -- whether biological or neuromorphic.

\section{Discussion}

\paragraph{Biological plausibility}
The main thrust behind our method is to provide a systematic approximation to spatiotemporal learning that is local both in space and time, thus respecting physical and biological constraints.
Approaching the problem as the minimization of a local energy functional ensures spatial locality. 
The outcome of variational optimization is easy to compute using elementary calculus of variations, but requires information about future errors that is non-local-in-time. 
This is to be expected, as true optimization of a temporal task requires perfect knowledge of future states.
More specifically, in order to preserve causality, the network has to learn to predict the discounted future error.
Because of its exponential weighting, the discounted future error is mostly sensitive to the near future, making it feasible to approximate it using local-in-time quantities.
Here, we use the look-ahead operator as a first order approximation, which is also the basis of the \gle framework.
In this work we go beyond this, by learning backward weights to compensate for the approximated future errors.
This backward learning can be combined with methods developed to solve the problem of weight transport, most notably the \pal framework.
All these ingredients can be assembled into a cortical microcircuit that implements biologically plausible learning, and constitutes a blueprint for brain models and for neuromorphic applications.
In this way, we have a full framework that learns forward and backward weights to achieve a local spatiotemporal approximation of the \bptt algorithm.

\paragraph{Comparison to previous results}

We start by considering \am, \bptt, and \rtrl. 
These methods provide the exact gradients for learning spatiotemporal tasks, but have major shortcomings as models of biological processes. 
\bptt (and, equivalently, \am) requires perfect knowledge of the future states of the network, which clearly violates causality.
In contrast, while \rtrl conserves causality, it needs to store in-memory all past influences of all synaptic weights on all neuronal activities, in the so-called influence tensor. 
Each element of this tensor connects a neuron with some far-away synapse, which effectively violates spatial locality. 
Furthermore, going beyond biology, the resulting cubic scaling of this method makes it highly impractical for large-scale networks. 
Approximations to \rtrl, such as \rflo~\cite{marschall2020unified}, e-prop \cite{murray2019local,bellec2020solution}, FORCE~\cite{sussillo2009generating} and FOLLOW~\cite{gilra2017predicting}, alleviate this issue by neglecting non-local terms in the influence tensor. 
This can lead to more biologically plausible implementations, but only at the cost of not being able to propagate temporal gradients more than one layer deep (or, equivalently, one neuron deep in recurrent networks).
\vle addresses these challenges directly, being a local, causal approximation of \bptt.

This work builds upon and combines the insights stemming from several previous publications, especially (Generalized) Latent Equilibrium~\cite{haider2021latent,ellenberger2025backpropagation} and Neuronal Least Action~\cite{senn2024neuronal}. 
When considering \gle, our method is an extension in two senses.
First, it presents \gle as the consequence of a variational optimization procedure followed by a linear approximation for the discounted future error.
That is to say, \vle takes inspiration from physics and uses a global minimization principle as a starting point. 
This allows an elementary derivation of the equations governing spatiotemporal learning in systems of neurons, a much simpler analytical derivation of a result that is equivalent to the well-known \am.
Physically speaking, the energy we are minimizing carries information about the local ``frustration'' or mismatch at the level of individual neurons, caused by the difference between the input a neuron receives from a previous layer and the input it is expecting to receive, as represented by its prospective membrane voltage.
This mismatch or error information is then transferred from the output layers upstream, via the error signals through error neurons.
Thus, the outcome of the method is compatible with \gle also at the level of specific microcircuit implementations.
Nonetheless, the variational derivation gives firmer grounding to the whole method, at least in terms of technical accessibility.
Secondly, \vle improves on \gle by correcting for this approximation, training the backward weights during learning.
This has the apparent disadvantage of doubling the number of trainable parameters, but can be important in situations where the linear approximations to the discounted future error can potentially break down.
This is most notably the case when there are multiple frequency components in a signal for which the erroneous gain modulation of \gle causes distorted responses.
Note also that in order to achieve fully bio-plausible learning the backward weights should be learned in any case, by e.g. the \pal algorithm. 
Thus, for a fully biologically plausible implementation of \bptt, the backward learning rule of \vle becomes a necessity.

Our model is similar to \nla in taking inspiration from physics but diverges from it in important ways.
The most crucial difference is that \nla cannot perform temporal credit assignment. 
Indeed, other than imposing a low-pass filter on its inputs, an \nla network effectively reacts instantaneously to external stimuli, and can neither carry out nor learn temporal sequence processing.
In contrast, in \vle, prospectivity is not only taking place at the level of the forward ``inference'' neurons, but neurons in the backward error path can also be prospective, to reduce the lag caused by the membrane integration.
Overall, there are two time constants that can be leveraged to perform rich temporal processing, one associated with membrane integration, and a second governing prospective output rates.
This central difference is what enables temporal processing, and is the key to approximate the \am and \bptt.
Additionally, the implicit microcircuit implementation behind \vle is much more symmetric than the one from \nla, where errors are transmitted through the use of instantaneous inter-neurons. 
In our microcircuits, we make use of the same type of neurons for inference and errors, connected in a symmetric way (see \cref{fig:circuit}a).

Finally,  it is important to emphasize that some of the assumptions in \vle, such as the strict connectivity scheme imposed by the derivation, can be substantially loosened to account for the great diversity found in biology.
For this, we can take inspiration from the microcircuit implementations recently developed by \cite{max2025backpropagation}, which show how methods such as \gle can be implemented in biologically plausible microcircuits.

\paragraph{Conclusions}

In this work we present \vle, a systematic approximation of the \am and \bptt that is local both in space and time.
It thus lends itself both as a model of cortical computation, and as a framework for implementing \bptt within brain-inspired neuromorphic hardware. 
Our work improves on previous results by better approximating temporal error transmission in the backward pathway.
Summarizing the above, \vle takes two important steps beyond \gle, one theoretical and one algorithmic. 
The new theory is more rigorously grounded and allows a clear assessment of any approximations to the exact gradient. 
The algorithmic innovation involves learning the backward weights $B$. 
By learning these backward weights we can correct for the gain error induced by the enforced temporal locality of the error dynamics. 
This leads to a better approximation of \am, as well as generally more stable dynamics, as our experiments show. 
Our results also show a significantly higher robustness of \vle with respect to high-frequency noise, which adds an essential ingredient to the model’s plausibility in analog physics -- whether biological or neuromorphic.
Thus, for spatiotemporal tasks that require the processing of signals with a diverse frequency range, \vle enables enhanced stability and more accurate learning. 
In combination with \pal, the backward learning in \vle allows for robust, biologically plausible (fully local) spatiotemporal learning in physical networks.

\section*{Acknowledgements}
We gratefully acknowledge funding from the European Union under grant agreement \#101147319 (EBRAINS 2.0). We would like to express particular gratitude for the ongoing support from the Manfred Stärk Foundation.

\section*{Author Contributions}
F.B., S.B., W.S., and M.A.P. conceived the core ideas and designed the project. P.H. developed and maintained the GLE codebase, which served as the basis for the VLE codebase, developed by S.B. S.B. performed the simulations.  All authors contributed to the development of the theory.  F.B. and S.B. wrote a first version of the manuscript. All authors contributed to the final manuscript.

\printbibliography{}


\appendix

\newpage
\FloatBarrier

\section{Supplemental Information}

\subsection{Theoretical derivation}
\paragraph{Detailed derivation of VLE equations}

The network of neurons is described by a global energy that consists of the sum of local error signals $e_i$ and a global cost $C$, \cref{total_energy}
\begin{equation}\notag
 E(t) = \frac12 \sum_i e_i^2(t) + \beta C(t)
\end{equation}

The membrane potential of a neuron follows the leaky-integrator dynamics
\begin{align}
    \taum_i\dot{u}_i = -u_i + e_i + \sum_j W_{ij} \varphi_j(\ubrever_j) + b_i \,,
    \label{eqn:met_udot}
\end{align}
where each neuron integrates bottom-up input $\varphi_j(\ubrever_j)$ via synaptic weights $W_{ij}$ with bias $b_i$ and top-down errors $e_i$.
The output rate of lower layer neurons is defined by the activation $\varphi_j$ as a function of the prospective voltage $\ubrever_j=u_j+\taur_j\dot{u}_j$ that is a look-ahead with time constant $\taur_j$.
Each neuron $i$ will thus low-pass filter its inputs with its membrane time constant $\taum_i$.
This low-pass filter induces a lag of the order of $\taum_i$ that is partly reversed by the prospective output $\varphi_i(\ubrever_i)$ which implements a temporal advance in the order of $\taur_i$.

Rearranging \cref{eqn:met_udot} and using $\ubrevem_i=u_i+\taum_i\dot{u}_i$ yields the definition of the local mismatch errors
\begin{align}
    e_i = \ubrevem_i - \sum_j W_{ij} \varphi_j(\ubrever_j) - b_i\,.
    \label{eqn:met_edef}
\end{align} 
The local mismatch error can be seen as the difference between what the neuron thinks it should be doing in the near future ($\ubrevem_i$) and what its predecessors think it should be doing.
To minimize this local mismatch error alongside the global cost, we use a variational approach and minimize the integrated energy $\intE = \int_{t_1}^{t_2} \dd t \, E(t)$.
Here, we use the integral over time to account for the general case of temporal processing tasks, which require to minimize the integrated cost $\intC = \int_{t_1}^{t_2} \dd t \, C(t)$ (see next section for details on this).

To find the minima of such a functional $\intE$, a variational approach is well-suited.
We consider small variations $\alpha\eta(t)$ of the membrane voltage such that $u_\alpha(t)=u+\alpha\eta(t)$ with $\eta(t)$ being an arbitrary function that vanishes at the boundaries.
Importantly, this variation of $u$ subsequently affects its temporal derivative such that we also get $\dot{u}_\alpha(t) = \dot{u} + \alpha\dot{\eta}(t)$.

To consider variations of $u$ in $E$, we note that with \cref{eqn:met_edef} and assuming the cost $C=C(u(t), \dot{u}(t))$ to be a function of the network output, we can write the energy as a function of the voltage $u$ and its temporal derivative $\dot{u}$, namely $E(t)=E(u(t), \dot{u}(t))$.
According to the calculus of variations, the integrated energy $\intE_\alpha = \int_{t_1}^{t_2} E(u_\alpha(t), \dot{u}_\alpha(t))\dd t$ has a minimum for $\alpha=0$ and we can write
\begin{align}
    \left.\dfrac{\dd \intE_\alpha}{\dd \alpha} \right|_{\alpha=0} &= \int_{t_1}^{t_2}\left.\dfrac{\dd E(u_\alpha(t), \dot{u}_\alpha(t))}{\dd \alpha} \right|_{\alpha=0}\dd t\,.
\end{align}

The result yields the Euler-Lagrange equations
\begin{align}
    0 = \frac{\partial E}{\partial u_i} - \frac{\dd}{\dd t} \frac{\partial E}{\partial \udot_i}\,,
\end{align}
which we can now use to derive error-backpropagation in the neuronal dynamics:
\begin{align}
	\begin{array}{ccccc}
		\multicolumn{2}{c}{\frac{\partial E}{\partial u_i}} &=& \multicolumn{2}{c}{\frac{\dd}{\dd t} \frac{\partial E}{\partial \udot_i}} \\
		\multicolumn{2}{c}{\frac{\partial E}{\partial \ubrevem_i} \frac{\partial \ubrevem_i}{\partial u_i} + \frac{\partial E}{\partial \ubrever_i} \frac{\partial \ubrever_i}{\partial u_i}} &=& \multicolumn{2}{c}{\frac{\dd}{\dd t} \frac{\partial E}{\partial \ubrevem_i} \frac{\partial \ubrevem_i}{\partial \udot_i} + \frac{\dd}{\dd t} \frac{\partial E}{\partial \ubrever_i} \frac{\partial \ubrever_i}{\partial \udot_i}} \\
		{\scriptstyle (1- \taum_i \frac{\dd}{\dd t})} & \frac{\partial E}{\partial \ubrevem_i} &=& {\scriptstyle (1- \taur_i \frac{\dd}{\dd t})} & \roundpar{-\frac{\partial E}{\partial \ubrever_i}} \\
		{\scriptstyle (1- \taum_i \frac{\dd}{\dd t})} & \overbrace{e_i} &=& {\scriptstyle (1- \taur_i \frac{\dd}{\dd t})} & \overbrace{\scriptstyle \varphi'_i(\ubrever_i) \sum_j  W_{ji} e_j} \\
		 & \widehat{e_i}^{\mathrm m} &=&  & \reallywidehat{\scriptstyle \varphi'_i(\breve{u}_i^r) \sum_j  W_{ji} e_j}^{\mathrm r}\,.
		\label{eqn:met_eulerlagrangetoam}
	\end{array}
\end{align}

Next, we apply the future-discount operator to both sides of the equation to get the future-discounted error:
\begin{align}
    e_i  = \tilde\varepsilon_i^\mathrm{m} \; \textrm{with}\; \;  \varepsilon_i = {\reallywidehat{\varphi'_i \sum_k W_{ki} e_k}}^\mathrm{r}\,.
    \label{eqn:met_eproptle}
\end{align}

This is the same result we get from the \am.
Because the future-discounted error cannot be obtained without knowledge about the future, it is biologically not plausible.
As a biologically plausible approximation, we use the prospective error instead.
We further approximate \cref{eqn:met_eproptle} towards biological plausibility by using, instead of the look-back operator, a low-pass filter that corresponds to the membrane integration of neurons.
The resulting error dynamics
\begin{align}
    e_i  = \brevem{\epsilon}_i \; \textrm{with}\; \;  \epsilon_i = {\barr{\varphi'_i \sum_k W_{ki} e_k}}\,
\end{align}
recover the equations of \gle.
The approximated error dynamics are similar to the neuron dynamics of the forward path, as they low-pass filter input and use prospective output rates.
An implementation into a neuronal circuit thus uses the same neuron model for forward and error neurons and yields a symmetric circuit as shown in \cref{fig:circuit}a.

Note that we omitted the external cost here, which only affects output neurons and is assumed to be a function of the output rates $C=C(\ubrever(t))$.
Because output neurons do not project onto other neurons, i.e. $W^{\mathrm{out}}_{ij}=0$, their error can be written as
\begin{align}
    \epsilon_{i, \rm{out}} = - \beta \barr{ \dfrac{\partial C(\ubrever)}{\partial \ubrever_{i, \rm{out}}}}\,.
\end{align}

For supervised learning, a typical cost to minimize is the mean squared error $C(t)=\frac{1}{2}(r^{\rm trg} - r^{\rm out})^2$, for which the target error of the output layer is given by
\begin{align}
    \epsilon_{i, \rm{out}} = \beta \barr{\varphi'(\ubrever)\,(r^{\rm trg} - \varphi(\ubrever))}\,.
\end{align}

\paragraph{Minimizing the energy and minimizing the cost}

In \vle, the ultimate objective for the network is to minimize the cost $\intC$, of which the integrated energy $\intE$ is but a proxy.
Taking inspiration from \cite{scellier2017equilibrium}, we show how to recover cost minimization in the limit of small $\beta$.
In order to show how this works explicitly, we go back to \cref{total_energy}: $E(t) = \frac12 \sum_i e_i^2(t) + \beta C(t)$

Being a sum of quadratic terms, it is easy to see that in the case of zero nudging, $\beta = 0$, the trivial solution to $\delta \intE = 0$ is $e_i = 0$ $\forall i$.
By the same token, for small nudging strength $\beta$, the errors are also small, $\bm e = o(\beta)$.
Because $E$ sums up the squared errors, the energy for small $\beta$ is dominated by the cost term, $E(t) = \beta  C(t) +  o(\beta^2)$.
Given that $\intE = \int_{t_1}^{t_2} E(t) \,\dd t$, we have $\lim_{\beta\to 0} \frac{1}{\beta} \intE = \intC$.
Taking the total derivative with respect to a network parameter $\bm \theta$ on both sides yields
\begin{equation}
      \lim_{\beta\to 0} \frac{1}{\beta} \frac{\dd \intE}{\dd \bm \theta} = \frac{\dd \intC}{\dd\bm \theta} \;.
\end{equation} 
In order to see how this leads to gradient descent cost minimization, we explicitly calculate this total derivative of the functional $\intE[\bm u(\theta)]$
\begin{equation}
\frac{\dd \intE}{\dd \bm \theta} = 
\int_{t_1}^{t_2} \dd t \, \left( \delta_u \intE \frac{\dd \bm{u} }{\dd \bm \theta} \right)+ \frac{\partial \intE}{\partial \bm \theta} = \frac{\partial \intE}{\partial \bm \theta} \;,
\label{eq:dEdtheta}
\end{equation}
where we used the Gateaux derivative for functional differentiation, and we are considering the derivative evaluated at the optimized $u$ trajectory (so-called on-shell). Since the global and partial derivatives of $\intE$ with respect to $\bm \theta$ are equal, we get $\lim_{\beta\to 0} \frac{1}{\beta} \frac{\partial \intE}{\partial \bm \theta} = \frac{\dd \intC}{\dd\bm \theta} $.
This means that minimizing the integrated energy with respect to the network parameters is equivalent to minimizing the integrated cost.

Notice that this on-shell calculation assumes compact variations, which vanish at the boundaries. This assumption can become incorrect when considering large variations of network parameters from beginning to end of training. Our result is only exact within an adiabatic approximation, which considers slowly changing parameters that can be treated as constant functions. This represents a further, albeit common, approximation. With this, the learning rule \cref{plasticity} then becomes approximately cost-gradient in the limit of small nudging,
\begin{equation}
       \lim_{\beta\to 0} \frac{1}{\beta} {\bm{\dot \theta}}  \propto - \lim_{\beta\to 0} \frac{1}{\beta} \frac{\partial E}{\partial \bm \theta} \propto - \frac{\dd C}{\dd\bm \theta} \;.
\end{equation}

In the specific case of learning the forward weights, the partial derivative $\frac{\partial E}{\partial \bm W}$ can be easily calculated since it is a local quantity, and in fact involves the local error in the apical dendrite.
This way, the learning rule \cref{Wlearning} becomes approximately cost-gradient in the limit of small nudging.

\paragraph{Learning the backward weights}

We introduce backward weights $B_{ik}$  
 \begin{align}
    e_i & = \breve{\epsilon}_i^\mathrm{m} \; & \textrm{with}\; \;  \epsilon_i & = {\overline{\varphi'_i \sum_k B_{ik} e_k}}^{\;\mathrm{r}} \;.
\end{align}
The backward weights can learn to mirror their reciprocal forward synapses $W_{ji}$ through specific learning rules, e.g., \pal \cite{max2024learning} 
\begin{align}
	\ep_{ij} &= - \xi_i^\TT B_{ij} \hat r_j + \frac\alpha2 \roundpar{B_{ij}}^2 \\
	\Rightarrow \quad \dot B_{ij} &\sim - \frac{\partial \ep_{ij}}{\partial B_{ij}} = \xi_i \hat r_j^\TT - \alpha B_{ij},
\end{align}
where $\xi_i$ is a noise signal, the autocorrelation properties of which can be used to learn the correct backward weights.

\vle uses this kind of mechanism to also deal with the causality (temporal) issue. By learning backward weights that compensate the gain offset in \gle, the network can better converge to the exact \am solution, while preserving spatial and temporal locality. 

To see how this works, we start by defining instantaneous errors for \am, $\epsilon_i = \varphi'_i(\ubrever_i) \sum_j W_{ji} e_j$, and \vle, $\epsilon_i = \varphi'_i(\ubrever_i) \sum_j B_{ij} e_j$, to write down how they propagate:
\begin{align}
	\text{\am}: e_i & = \tilde\varepsilon_i^\mathrm{m} \; & \textrm{with}\; \;  \varepsilon_i& = {\reallywidehat{\varphi'_i \sum_j W_{ji} e_j}}^\mathrm{r} \\
	\text{\vle}: e_i & = \breve{\epsilon}_i^\mathrm{m} \; & \textrm{with}\; \;  \epsilon_i & = {\overline{\varphi'_i \sum_j B_{ij} e_j}}^{\;\mathrm{r}}.
\end{align}
To equalize the two, we can learn a $\bm B$ that compensates for the difference between the look-ahead and the discounted-future operators, so that $e_i^\text{\am} \mustbe e_i^\text{\vle}$. In fact we can consider the even stronger constraint:
\begin{equation}
	\sum_j W_{ji} \tilde\varepsilon_j^\mathrm{m} \mustbe \sum_j B_{ij} \breve{\epsilon}_j^\mathrm{m}.
	\label{eqn:optimalbcomponents}
\end{equation}

To achieve the above, we could simply use local gradient descent to minimize $\eb_{i} = \frac12 \roundpar{\sum_j W_{ji}\tilde\varepsilon_j^\mathrm{m} - \sum_j B_{ij} \breve{\epsilon}_j^\mathrm{m}}^2$, obtaining $\dot B_{ij} = \roundpar{\sum_k W_{ki} \tilde\varepsilon_k^\mathrm{m} - \sum_k B_{ik} \breve{\epsilon}_k^\mathrm{m}} \breve{\epsilon}_j^\mathrm{m}$. Unfortunately, such a learning rule still uses the discounted future $\tildem{x}$, which is unavailable to the network in real time. A bio-plausible solution can be found by applying the look-back operator to both sides of this learning rule. Namely, we can define the loss
\begin{align}
	\eg_{i} &= \frac12 \roundpar{\sum_k W_{ki}  \bhrr{\epsilon}{k} - \sum_k B_{ik} \bhmm{\epsilon}{k}}^2\\
	\Rightarrow \quad \dot B_{ij} &\sim - \frac{\partial \eg_{i}}{\partial B_{ij}} = \roundpar{\sum_k W_{ki}  \bhrr{\epsilon}{k} - \sum_k B_{ik} \bhmm{\epsilon}{k}} \bhmm{\epsilon}{j};
    \label{eqn:met_B_dot}
\end{align}
which only includes quantities available on-line, since both look-aheads and look-backs are causal.
We used $\varepsilon_j = \bhrr{\epsilon}{j}$ to compensate for the difference of the low-pass filter in \vle and the look-back operator in \am.

To see that $\eg$ has the same minima as $\eb$ we simply apply the look-back operator to its argument (\cref{eqn:optimalbcomponents}), which conserves the roots of $\bm{\dot{B}}$, i.e., the minima of $\eb$.

This proposal still suffers from the weight transport problem, as is made evident by the presence of $W_{ji}$ in the learning rule for $B_{ij}$. To solve this we proceed in two steps.
First, neurons learn to align to the forward weights by harnessing intrinsic neuronal noise $\xi_i$, as proposed in \cite{max2024learning}: $\dot B_{ik}^{*} \sim \xi_i \hat r_k - \alpha B_{ik}^{*}$.
Then, neurons would learn to compensate the \gle gain distortion by minimizing the loss
\begin{align}
    \eg_{i} &= \frac12 \roundpar{\sum_k B_{ik}^*  \bhrr{\epsilon}{k} - \sum_k B_{ik} \bhmm{\epsilon}{k}}^2 \; .
    \label{eq:galenergy}
\end{align}
The implementation of \pal via \cref{eq:galenergy} requires the use of larger microcircuits, which have already been deployed successfully by \cite{max2024learning}, and are thus not the focus of this work.
Assuming that \pal works as designed, we only use \cref{eqn:gal} in our simulations for simplicity.

\subsection{Numerical implementation}
We use forward Euler integration to solve the set of coupled differential equations that define the dynamics of the network of neurons.
While all computations work with vector and matrix calculation, we use neuron indices here to highlight locality.
The membrane voltage of a neuron integrates the bottom-up input and top down error
\begin{align}
    \Delta u_i(t) = \dot{u}_i(t) = \frac{1}{\taum_i}(-u_i(t) + \sum_j W_{ij}r_j(t) + b_i + e_i(t))\,.
\end{align}

We then use finite differences to approximate ${\Delta u_i(t) \approx \frac{u_i(t+\dd t) - u_i(t)}{\dd t}}$.
The voltage at $t+\dd t$ can then be computed by
\begin{align}
    u_i(t+\dd t) = u_i(t) + \dd t \Delta u_i(t)\,.
\end{align}

The error neurons low-pass filter the top-down error and the derivative of their dynamics can be expressed similarly as
\begin{align}
    \Delta\epsilon_i(t)=\dot\epsilon_i(t) = \frac{1}{\taur_i}(-\epsilon_i(t) + \varphi'_i(\ubrever_i(t))\sum_j B_{ij}e_j(t)),
    \label{eqn:met_epsilon_int}
\end{align}
and the error neuron membrane potential gets updated via
\begin{align}
    \epsilon_i(t + \dd t) = \epsilon_i(t) + \dd t \Delta\epsilon_i(t)\,.
\end{align}

The learning of the parameters follows the same scheme and in the case of the weights becomes
\begin{align}
    W_{ij}(t+\dd t) = W_{ij}(t) + \eta_W e_i(t)r_j(t),
\end{align} 
with $\eta_W=\dfrac{\dd t}{\tau^W}$ defined by the learning rate for $W$.

To learn the backward weights according to \cref{eqn:met_B_dot}, we need to compute the look-ahead and look-back operator.
Applying both operators sequentially to the error variable, i.e. $\bhrr{\epsilon}{i} = \epsilon_i - (\taur_i)^2\ddot{\epsilon}_i$, requires second order temporal derivatives.
The update rule for $\bm B$ then becomes
\begin{equation}
\begin{aligned}
    B_{ij}(t+\dd t) = B_{ij}(t) + \eta_B \roundpar{\sum_k W_{ki}\bhrr{\epsilon}{k}(t) - B_{ik}\bhmm{\epsilon}{k}(t)}\bhmm{\epsilon}{j}(t)\,,
\end{aligned}
\end{equation}
with $\eta_B=\dfrac{\dd t}{\tau^{\rm B}}$ again being a learning rate and
\begin{align}
    \bhrr{\epsilon}{j}(t) = \epsilon_j(t) - (\taur_j)^2\dfrac{\epsilon_j(t+\dd t) - 2\epsilon_j(t) + \epsilon_j(t-\dd t)}{\dd t^2}\,,
\end{align}
using the instantaneous input integrated in \cref{eqn:met_epsilon_int} to get $\epsilon(t+\dd t)$.
The second order derivative can also be expressed as finite difference of first order differentials  $\ddot{\epsilon}_j(t)=\dfrac{\dot{\epsilon}_j(t) - \dot{\epsilon}_j(t - \dd t)}{\dd t}$.
While we obtain $\dot{\epsilon}_j(t)$ by rearranging \cref{eqn:met_epsilon_int}, our implementation needs to store one additional past value to compute the second order derivative. 
We note that such workarounds when estimating temporal derivatives are a common issue when discretizing differential equations that are meant to be continuous.
\Cref{alg:VLE}, adapted from~\cite{ellenberger2025backpropagation}, summarizes the system of equations as simulated in our experiments.

\begin{algorithm}
  \small
  \caption{Forward Euler simulation of VLE network}
  \begin{algorithmic}
    \State initialize network parameters $\bm \theta = \{\bm W, \bm B, \bm b, \Taum, \Taur\}$
    \State initialize network states at $t=0$: $\bm u(0)$, $\bm \epsilon(0)$, $\bm r(0)$, $\bm e(0)$
    \For{time step $t$ in $[0, T]$ with step size $\dd t$}
      \For{layer $\ell$ from $1$ to $L$}
        \If{$\ell = L$}
        \State calculate instantaneous target error:
        \State $\bs{e}^{\mathrm{inst}}_{L}(t) \gets \beta\bs{\varphi}'_{L}(t) \odot (\bs{r}^{\mathrm{trg}}(t) - \bs{r}_{L}(t))$
        \Else
        \State propagate feedback error signals:
        \State $\bs{e}^{\mathrm{inst}}_{\ell}(t) \gets \bs{\varphi}'_{\ell}(t) \odot \bs{B}_{\ell+1}(t) \bs{e}_{\ell+1}(t)$
        \EndIf
        \State sum input currents:
        \State $\quad \bs I_\ell(t) = \bs{W}_{\ell}(t) \bs{r}_{\ell-1}(t) + \bs{b}_{\ell}(t) + \gamma\bs{e}_{\ell}(t)$
        \State approximate membrane potential derivatives:
        \State $\quad\Delta \bs{u}_{\ell}(t)  \gets {(\btaum_{\ell}(t))}^{-1} \odot \left(-\bs{u}_{\ell}(t) + \bs I_{\ell}(t)\right)$
        \State update membrane potentials:
        \State $\quad\bs{u}_{\ell}(t + \dd t) \gets \bs{u}_{\ell}(t) + \dd t \Delta \bs{u}_{\ell}(t)$
        \State approximate error potential derivatives:
        \State $\quad\Delta \bs{\epsilon}_{\ell}(t) \gets {(\btaur_{\ell}(t))}^{-1} \odot \left(-\bs{\epsilon}_{\ell}(t) + \bs{e}^{\mathrm{inst}}_{\ell}(t)\right)$
        \State update error potentials:
        \State $\quad\bs{\epsilon}_{\ell}(t + \dd t) \gets \bs{\epsilon}_{\ell}(t) + \dd t \Delta \bs{\epsilon}_{\ell}(t)$
        \State calculate prospective errors:
        \State $\quad\bs{e}_{\ell}(t + \dd t) \gets \bs{\epsilon}_{\ell}(t) + \btaum_{\ell}(t) \odot \Delta \bs{\epsilon}_{\ell}(t)$
        \State calculate prospective outputs:
        \State $\quad\bs{r}_{\ell}(t +\dd t) \gets \bs{\varphi}\left(\bs{u}_{\ell}(t) + \btaur_{\ell}(t) \odot \Delta \bs{u}_{\ell}(t)\right)$
        \State update synaptic weights:
        \State $\quad\bs{W}_{\ell}(t +\dd t) \gets \bs{W}_{\ell}(t) + \eta_{W} \bs{e}_{\ell}(t) \bs{r}^{\TT}_{\ell-1}(t)$
        \State update backward weights:
        \State $\quad\Delta^2 \bs{\epsilon}_\ell(t) \gets (\Delta\bs{\epsilon}_\ell(t) - \Delta\bs{\epsilon}_\ell(t-\dd t))/\dd t$
        \State $\quad\bhmm{\bs{\epsilon}}{\ell} (t) \gets \bs{\epsilon}_\ell(t) - (\btaum_\ell)^2(t) \odot \Delta^2 \bs{\epsilon}_\ell(t)$
        \State $\quad\bhrr{\bs{\epsilon}}{\ell} (t) \gets \bs{\epsilon}_\ell(t) - (\btaur_\ell)^2(t) \odot \Delta^2 \bs{\epsilon}_\ell(t)$
        \State $\quad\bs{B}_{\ell}(t +\dd t) \gets \bs{B}_{\ell}(t) + \eta_{B} (\bs{W}_{\ell}^\TT \bhrr{\bs{\epsilon}}{\ell}(t) - \bs{B}_{\ell} \bhmm{\bs{\epsilon}}{\ell}(t)) \bhmm{\bs{\epsilon}}{\ell}(t)^\TT$
        \State update biases:
        \State $\quad\bs{b}_{\ell}(t + \dd t) \gets \bs{b}_{\ell}(t) + \eta_{b} \bs{e}_{\ell}(t)$
        \State update membrane time constants:
        \State $\quad\bs{\tau}_\ell^{\mathrm{m}}(t + \dd t) \gets \bs{\tau}_\ell^{\mathrm{m}}(t) - \eta_{\tau} \bs{e}_{\ell}(t) \odot \Delta \bs{u}_{\ell}(t)$
        \State update prospective time constants:
        \State $\quad\bs{\tau}_\ell^{\mathrm{r}}(t +\dd t) \gets \bs{\tau}_\ell^{\mathrm{r}}(t) + \eta_{\tau} \bs{e}^{\mathrm{inst}}_{\ell}(t) \odot \Delta \bs{u}_{\ell}(t)$
      \EndFor
    \EndFor
  \end{algorithmic}
  \label{alg:VLE}
\end{algorithm}

\subsection{Experimental setup}

\tb{Lagline}
In this experiment, we set up a simple case of small-network learning to test how a student network learns the output of a teacher network with the same topology.
Because of the simplicity and the choice of nonlinearities, the task has a unique solution: learning the teacher weights.
The time constants of the neurons are $\taum_0=0.4$, $\taum_1=0.2$, $\taur=\dd t$, with a softplus nonlinearity and the teacher weights $W^*_0=1$ and $W^*_1=2$ with no biases.
The weights of the student are initialized to the opposite of the teacher weights $W_0=-1$ and $W_1=-2$ and biases are not learned.
We investigate three scenarios: learning the backward weights $\bm B$, using $\bm{B}=\bm{W}^T$ and constant $\bm{B}=-\bm{W}^{*T}$.
The learning rates are $\eta_W=0.01$ and $\eta_B=0.1$, nudged with $\gamma=1$ at $\beta=0.5$, and the simulation time step $\dd t=0.01$.
The input is a sine with frequency $f=$~\SI{1}{Hz}.
We initialize the network for $T_{\rm init}=$~\SI{10}{s}, train for $T_{\rm train}=$~\SI{1000}{s} and evaluate for $T_{\rm test}=$~\SI{10}{s}.
We use a mean squared error loss and all parameters were optimized using an \texttt{SGD} optimizer.
The experiment was run locally on a laptop using the CPU only.

\tb{Lagnet}
The network used in this experiment consists of one neuron in the input layer, two hidden layers with two and three neurons and three output neurons, each with $\tanh$ activations and no biases.
The input signal consists of multiple overlaid sines with $N$ frequency components $f_i\in\{0.44, 0.55, 0.77, 1.3\}$, amplitudes $a_i\in\{0.4, 0.3, 0.2, 0.2\}$ and randomly sampled shifts $s_i \sim \mathcal{U}[0, 1/f_i]$.
We normalize the amplitudes of the components $A_{i}=\dfrac{a_i}{(\sum_j a_j^2)^{1/2}}$, such that the input signal becomes ${r_0(t) = \sum_i A_{i} \sin(2\pi f_i (t-s_i))}$.
The time constants of the networks are $\{0.3, 0.7\}$ for the first layer, $\{0.1, 0.5, 0.9\}$ for the second layer, $\{0.2, 0.4, 0.8\}$ for the last layer and $\taur=0.05$ for all layers.
Teacher weights are drawn from a Xavier uniform distribution $\mathcal{U}[-\sqrt{6 / (n_\mathrm{in} + n_\mathrm{out})}, \sqrt{6 / (n_\mathrm{in} + n_\mathrm{out})}]$ and multiplied by $\frac1N\sum_k A_{k}\sqrt{1 + (2\pi f_k\tau^\mathrm{m}_i)^2}$ to compensate the gain decrease from the low-pass filtering.
Student weights are drawn from the same distribution without compensation.
The hyperparameters are $\eta_W=0.003$ and $\eta_B=3$ for learning $B$ and $\eta_W=0.01$ for $B=W^T$, nudged with $\gamma=1$ at $\beta=0.5$.
To avoid numerical instabilities that arise from estimating the second order derivative in the $B$ learning, $B$ gradients are clipped to 0.01.
The performance of the trained model is evaluated on a batch of 32 test signals with 10 frequency components each, sampled uniformly in the range of the training signal between \SIrange{0.44}{1.3}{Hz}.
The network was initialized for $T_{\rm init}=$~\SI{10}{s}, trained for $T_{\rm train}=$~\SI{10000}{s} and evaluated for $T_{\rm eval}=$~\SI{10}{s} and the simulation time step is $\dd t=0.01$.
All parameters were optimized using an \texttt{SGD} optimizer and a mean squared error loss and hyperparameters were optimized with a grid search.
The experiment and hyperparameter optimization were run locally on a laptop using the CPU only.

\tb{Lagnet with noise}
This experiment follows the setup of the lagnet experiment, but uses noise on the student input, but not the teacher and consequently the target.
The noise is added during training, but not during testing.
The noise is Gaussian white noise that is normalized to unit variance and then scaled by a factor $\sigma$.
For every noise level, hyperparameters were optimized as in the lagnet experiment.

\tb{Cartpole}
In this task we demonstrate the ability of our method to act as a controller.
The task is to swing up a pole on a cart and keep it upright by applying a force to the cart as described in \cite{barto1983neuronlike}.
The cartpole setup used is a reimplementation of the \texttt{gymnasium} code \cite{towers2025gymnasium}.
Because our method is currently only implemented in a supervised setting, we use a model predictive controller (MPC) \cite{rawlings2017model} as a teacher.
Balancing a cartpole is very unstable and is a challenging task for our method that is developed for continuous time but implemented in discrete time steps.
To avoid numerical instabilities that arise from an input signal taking one $\dd t$ per layer to arrive at the output layer, we use a separated forward and backward pass for our model that assures that the output corresponds to the given input and target.
The three inputs of the network are the position $x$ of the cart and the angle of the pole represented by $\sin\theta$ and $\cos\theta$.
We use $\sin$ and $\cos$ to avoid jumps in the variable $\theta$ because of its circular nature.
The challenge of the task is to not use the derivatives $\dot{x}$ and $\dot{\theta}$, which are usually provided as additional input to models solving the task.
The missing derivatives force the network to intrinsically perform temporal processing and generate similar latents.

We use a network with two hidden layers and 128 neurons each and $\dd t=0.02$.
The first layer is inspired by a PID controller and consists of 1/3 slow neurons, 1/3 instantaneous neurons and 1/3 prospective neurons.
The slow neurons ($0.5\leq\taum\leq1$, $\taur=\dd t$) integrate input, the instantaneous neurons ($\taum=\taur=\dd t$) are proportional to the input and the prospective neurons ($\taum=\dd t$, $0.2\leq\taur\leq1$) provide the derivative.
Time constants that are not $\dd t$ are initialized linearly within the given range.
The activation functions of the two hidden layers are $\tanh$ and the output layer is linear.
Forward weights $W$ and backward weights $B$ are initialized with Xavier normal distributions and biases are initialized to $0$.
The training data generated with the MPC controller consists of 1000 episodes with a length of 300 steps of step size $\dd t=0.02$.
700 episodes are used for training, 150 for validation and 150 for testing.
The model is tested in an online fashion by letting it control a cartpole directly and measuring how long it survives by keeping the cart within the boundaries [-2.4, 2.4].
We report the upright rate of the model as the mean success rate of 50 test trials with a trial being successful if the model survived for 500 steps and the pole was upright in the end.
We average the performance over the seeds 1 to 10.
For each trial, the initial states of the cartpole $x$, $\dot{x}$, $\theta$ and $\dot{\theta}$ are initialized randomly from a uniform distribution in the range of -0.05 and 0.05, for the angle around its downward position.
To train the model we use the \texttt{adam} optimizer with a learning rate of $\eta=0.0001$ and train the model for 1000 epochs.
We compare two \vle models, one that learns the backward weights $B$ with $\eta_B=0.0001$ and another model that does not learn $B$ and corresponds to \gle with feedback alignment.
We set $\gamma=0$, $\beta=1$ and train with randomly shuffled batches of size 50.
The experiment was run locally and hyperparameter optimization on a cluster using CPU.

\tb{\mnistId}
This experiment uses the dataset described in \cite{greydanus2024scaling}, but interpolated to 1440 datapoints per sample and is adapted from~\cite{ellenberger2025backpropagation}.
With $\dd t=$~\SI{0.05}{s} we get \SI{72}{s} per sample and we use 4000 training and 1000 validation samples.
The target is the one-hot label of the sample at every time step.
The network consists of one input, six hidden layers of 53 neurons with $\tanh$ activation, and ten output neurons with softmax activation and all neurons have a bias.
In every hidden layer, the membrane time constants form three groups: 18 neurons with $\taum = \dd t$, 18 with $\taum =$~\SI{0.6}{s}, 17 with $\taum =$~\SI{1.2}{s} and all hidden neurons have $\taur = \dd t$.
The output neurons are instantaneous with $\taum = \taur = \dd t$.
Weights are drawn from a Xavier normal distribution and biases start at zero and all scenarios share the same initial forward weights.
$B$ weights are drawn from the same distribution but are scaled by $(\dfrac{\taur}{\taum})^2$.
For error propagation, we use $\beta=1$ and $\gamma=0$ and parameters are optimized with the \texttt{adam} optimizer and a cosine scheduler.
Learning rates were optimized by grid search, yielding $0.001\dd t$ for learning $B$, $0.002\dd t$ for $B=W^T$ and $0.01$ for \bptt.
While the first two are applied online at each time step, \bptt computes one update per batch.
For training, we use batches of 100 samples each that are reshuffled every epoch and trained for 150 epochs.
After every epoch, the network is tested on the validation data.


\end{document}